\documentclass[a4paper,prb,reqno,twocolumn,showpacs,footinbib]{revtex4-2}
\usepackage[utf8]{inputenc}
\usepackage{amsmath}
\usepackage{amsfonts}
\usepackage{dsfont} 
\usepackage{amssymb}
\usepackage{makeidx}
\usepackage{graphicx}
\usepackage{color}
\usepackage{mathptmx}
\usepackage{changes,enumitem,todonotes}
\usepackage{soul}
\usepackage[mathscr]{euscript}
\usepackage{tikz}
\usepackage[normalem]{ulem}
\usepackage[utf8]{inputenc}
\usepackage{amsmath}
\usepackage{amsfonts}
\usepackage{amssymb}
\usepackage{subcaption} 
\usepackage{hyperref}
\hypersetup{colorlinks=true,linkcolor=blue,urlcolor=blue,citecolor=blue} 
\usepackage{mathtools} 
\usepackage{soul,xcolor}

\begin{document}
\setstcolor{red}
\title{Revival of transport reciprocity via quantum interference
in asymmetric nonlinear devices}
\author{Rupak Bag and Dibyendu Roy}
\affiliation{Raman Research Institute, Bengaluru 560080, India}

\begin{abstract}
Structural asymmetry combined with optical nonlinearity often leads to nonreciprocal light transport. We explore the mechanism by which 1-photon interference effects can revive reciprocity in such nonlinear models. To this end, we study correlated 2-photon scattering where an artificial atom is asymmetrically (a) side-coupled to an infinite waveguide at two spatially separated points, and (b) direct-coupled to two semi-infinite waveguides.
The setup (a) gives robust reciprocal transport for the two photons. 
However, the setup (b) shows a transition from a nonreciprocal to a reciprocal regime by tuning the interference effect via an additional tunneling path for photons between the two waveguides.  

\end{abstract}
\maketitle

Quantum transport is a vibrant research field that has greatly enriched our understanding of coherence in mesoscopic systems \cite{MelloKumar2004}. 
Advancements in  microwave technology   have further boosted transport studies with waveguide quantum electrodynamics (WQED) \cite{RMP_DRoy2017,chang2014quantum,Lalumiere_PhysRevA2013,van2013photon, RMP_Russian2023,LiuHouck2017,zhang2023superconducting,PhysRevX_TopologicalWaveguideQED}.   In such  superconducting setups \cite{zhang2023superconducting,mirhosseini2018superconducting}, confinement of transverse propagating modes below sub-wavelength scales ensures strong light-matter coupling to engineered quantum impurities \cite{PhysRevA_ShenFan2007,ShenFanPRL2007,DRoy_OpticalDiode_PRB2010,PhysRevLett_BarangerSimulation2013} and high few-photon nonlinearity \cite{Astafiev10a,Hoi_Router_PhysRevLett2011,HoiPRL2013,vrajitoarea2022ultrastrong}.
Strongly correlated photons can coherently propagate along the transmission line, in which one can engineer multiple pathways to design interesting quantum interference effects.
In this context, we investigate the reciprocal and nonreciprocal transport behaviors of two correlated photons and how they are affected by interfering pathways in the waveguide. 

Passive magnetic-field-free rectification for quantum light was initially proposed in WQED systems by combining few-photon nonlinearities with structural asymmetry \cite{DRoy_OpticalDiode_PRB2010,roy2013cascaded,RoyPRA2017}. Later studies have further extended these ideas \cite{FratiniPRL2014,Dai_PhysRevA2015,YagiUda_PhysRevA2017}, and realized a microwave diode with two superconducting qubits side-coupled to a single waveguide \cite{Hamann2018}. A similar concept was explored earlier in thermal rectification studies of two-terminal quantum nonlinear devices \cite{Segal_ThermalRectifier_PRL2005} made of a two-level system directly coupled to two thermal baths, which yield asymmetric heat currents for forward and reverse temperature gradients. Later, \textcite{Segal_PRL_sufficient} demonstrated that, along with the different particle occupation statistics in the harmonic baths than in the saturable system, broken structural symmetry is indeed sufficient for getting the rectifying behavior in this kind of direct-coupled spin-boson systems. 

Following such vast literature, it has almost become convenient to assume that structural asymmetry and nonlinear interactions alone are enough to produce rectifying behavior in passive nonlinear devices.
In this letter, we witness the breakdown of such notions in WQED systems due to 1-photon interference effects arising from multiple pathways in the waveguide. To this end, we consider 2-photon scattering, where an artificial two-level atom is asymmetrically side-coupled to an infinite waveguide at two different points, yielding a giant atom configuration \cite{PhysRevA_GiantAtomfirstpaper2014}. A two-level atom is saturated by a single photon; thus, 2-photon scattering leads to nonlinear transport in such models. 
Over the past few years, WQED with giant atoms has probed novel regimes of light-matter interaction, where interference between different pathways reshapes the effective decay rate of single and collective atoms \cite{PhysRevLett_DecohFreeGiant,PhysRevLett_GiantBoundState}, and time-delayed feedback in the interference loops can lead to pronounced non-Markovian dynamics \cite{PhysRevA_GiantTimeDelay2017,NonMarkov_Giant2024}. Using {\it exact} treatment, we find that an asymmetrically coupled giant atom always gives reciprocal nonlinear transport.  
Further, we modify the direct-coupled setup in \cite{DRoy_OpticalDiode_PRB2010,Segal_PRL_sufficient} and access the regimes where reciprocity revives for correlated photon pairs. 
We consider an atom directly coupled to two semi-infinite waveguides, plus a direct tunneling path for photons between the two waveguides. Tuning the interference effects 
via tunneling, we show a transition from nonreciprocal to reciprocal transport regimes in this generalized direct-coupled configuration.

\begin{figure}[h]
    \centering \includegraphics[width=8cm, height=2.3cm]{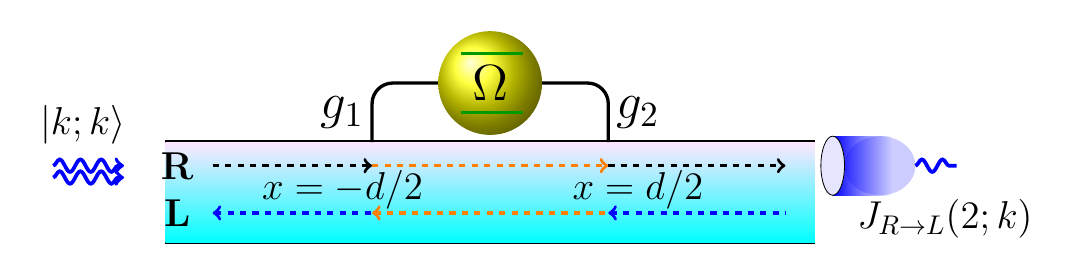}
    \caption{Side-coupled giant atom-waveguide setup. Two photons in the product state, with momentum $k$, are being injected into the right ($R$)-moving channel. A photo-detector measures the transmitted 2-photon current on the other side.  Here, $g_1\neq g_2$ induces a structural asymmetry in the system. }
    \label{Fig_GiantAtom}
\end{figure}

{1. \bf Giant atom-waveguide setup:} 
We start with a linear waveguide model that supports two channels with traveling modes towards the right ($R$) and left ($L$) directions (see Fig.~\ref{Fig_GiantAtom}).  
It is possible to design a meandered geometry of the waveguide on a superconducting chip which enables it to be coupled to an artificial atom at two spatially distinct locations \cite{kannan2020waveguide}. 
If the separation along the waveguide between the two coupling points is larger than the wavelength of the resonant photon associated with an atomic transition, the configuration becomes a giant atom–waveguide setup. The artificial atom can be modeled as an anharmonic bosonic site-$b$ (e.g., a transmon \cite{girvin2014circuit}) described by an effective Kerr Hamiltonian: $\hat{H}_{\text{b}}=\Omega \hat{b}^\dagger \hat{b}+U \hat{b}^\dagger\hat{b}^\dagger \hat{b}\hat{b}/2$ \cite{Longo_RadiationTrapping_2010PRL,DRoy_nonlinearPRA2012} with  $\hslash=1$. Here, $\Omega$ denotes the transition frequency between the ground and first excited state. 
$U$ characterizes the nonlinearity strength of the spectrum.   
We take the limit $U\to \infty$, where the anharmonic site-$b$ mimics the behavior of a two-level system.
The Hamiltonian of the full system is written as follows:
\begin{align}   \hat{H}^{g}=&\hat{H}_{\text{b}}+\hat{H}_{\text{w}}+ \int_{-\infty}^{\infty} dx~ g(x)\Big[\big(\hat{a}^\dagger_{R}(x)+\hat{a}^\dagger_{L}(x)\big)\hat{b}+\text{H.c.}\Big], \label{Eq_Hamiltonian_giantatom}
\end{align}
where $\hat{H}_{\text{w}}$ represents the Hamiltonian for an infinitely long waveguide, and it is given by 
\begin{align}
    &\hat{H}_{\text{w}}=-iv_g\int_{-\infty}^{\infty} dx \big[\hat{a}_R^\dagger(x)\partial_x \hat{a}_R(x)-\hat{a}_L^\dagger(x)\partial_x \hat{a}_L(x)\big].
\end{align}
We consider that the atom is side-coupled to the waveguide at points $x=-d/2$ and $x=d/2$  (see Fig.~\ref{Fig_GiantAtom}) with real coupling amplitudes $g_1$ and $g_2$. Here, we define $g(x)=g_1\delta(x+d/2)+g_2\delta(x-d/2)$.  
We employ the rotating-wave approximation for the light–matter interaction,  assuming $g_1,g_2 \ll \Omega$, and thereby restricting us to only resonant  processes in a narrow energy window. 
The real-space photon creation operators of the linear model \cite{RMP_DRoy2017} are denoted by $\hat{a}^\dagger_R(x)$ and $ \hat{a}^\dagger_L(x)$, respectively,  at position $x$ in the right ($R$)- and left ($L$)-moving channels. The group velocity of the linearized modes is denoted by $v_g$. We define the transmission current operator by $\hat{J}_{L\to R}(x)=v_g[\hat{a}^\dagger_R(x) \hat{a}_R(x)-\hat{a}^\dagger_L(x)\hat{a}_L(x)]$ for $x>d/2$ (see \cite{SM}), when photons are incident from the left of the atom. Our goal is to obtain the average 2-photon transmission current across the atom for $g_1\neq g_2$, breaking the parity of the setup.
Throughout the paper, we set $v_g=1$ and measure all energies in units of $v_g$.

 {\it 1-photon paradigm.}-- The transport properties of a single photon with momentum $k$ are solved exactly by taking an ansatz for the  scattering state of the Hamiltonian $\hat{H}^{g}$ within the single-excitation sector. The 1-photon transmission current from the left to the right of the  atom is given by $J_{L\to R}(1;k)=|t^{R}_k|^2/2\pi$, where the transmission amplitude is 
\begin{align}
    t^{R}_{k}=\frac{(k-\Omega-\Delta_k)}{( k-\Omega-\Delta_k)+i\Gamma_k/2}.\label{Eq_Giantatom_singlePhoton_transmission}
\end{align}
The Lamb shift $\Delta_k=2g_1g_2 \sin{kd},$ and the relaxation rate $\Gamma_k=2(g_1^2+g_2^2+2g_1g_2 \cos{kd})$ in Eq.~\ref{Eq_Giantatom_singlePhoton_transmission} show a strong dependence on momentum $(k)$, 
and these are symmetric in  $g_1,g_2$.  This indicates that the 1-photon transport is reciprocal across the giant atom. 
The structural asymmetry influences both the photonic wavefunction trapped within the giant-atom region in the waveguide (see \cite{SM}), and the phase associated with the atomic excitation amplitude $\phi^{R}_b(k)$, which is given by
\begin{align}
    \phi^{R}_b(k)=\frac{1}{\sqrt{2\pi}}\frac{g_1 e^{-ikd/2}+g_2 e^{ikd/2}}{( k-\Omega-\Delta_k)+i\Gamma_k/2}.
\end{align}
Therefore, due to interferences, the 1-photon excitation probability of the atom is the same for both left- and right-moving photons, i.e., $|\phi^{R}_b(k)|^2=|\phi^{L}_b(k)|^2$.
We perform an {\it exact} analysis in the 2-photon sector and show that the nonlinear part of the 2-photon ($L\to R$) transmission current depends only on the magnitude of $\phi^{R}_b(k)$, and not on its phase. 

  {\it 2-photon paradigm.}-- We use the Lippmann-Schwinger scattering theory for an {\it exact} treatment of 2-photon transport \cite{DRoy_Electrical_PRB2009,DRoy_nonlinearPRA2012}. Such a treatment is widely adopted to efficiently derive analytical expressions for 2-photon states in Markov WQED models \cite{PhysRevLett_BarangerGreens2013, Wenju_PhysRevA2023Markov}. In contrast, we employ this approach to obtain semi-analytical expressions for the nonlinear currents in the non-Markovian regime of giant atoms.
We separate the non-interacting and interacting parts of the 2-photon scattering states of $\hat{H}^{g}$ with energy $2k$. The interacting part involves
the expectations of 2-photon Green's function $\hat{G}_{o}^{+}(2k)=(2k-\hat{H}^{g}_o+i 0^{+})^{-1}$ that can be carried out semi-analytically using the roots $\{p_j\, |\forall j\in \mathbb{Z}^+\}$ of the self-energy equation $f_{+}(p)=p-\Omega-\Delta_{p}+ i\Gamma_{p}/2=0$.  Particularly, the expectation of $\hat{G}_{o}^{+}(2k)$ in the 2-photon computational basis state $|b;b\rangle$ at the impurity site-$b$, is symmetric in  $g_1, g_2$ (see \cite{SM}).
The 2-photon transmission  current is also decomposed into two components arising from non-interacting and interacting parts of the wavefunction as: $J_{L\to R}(2;k)={J}^{o}_{L\to R}(2;k)+\delta J_{L\to R}(2;k)$ \cite{DRoy_nonlinearPRA2012}. The  transmission current of two non-interacting photons is given by $J^{o}_{L\to R}(2;k)=2J_{L\to R}(1;k) \delta(0),$ where the Dirac delta function $\delta(k)$ is evaluated at $k=0$.  We further write the nonlinear  transmission current of two interacting photons as follows: $\delta J_{L\to R}(2;k)=\delta J^{c}_{L\to R}(2;k)+\delta J^{s}_{L\to R}(2;k)$, where the cross-correlation part $\delta J^{c}_{L\to R}(2;k)$ denotes a contribution from both interacting and non-interacting parts of the wavefunction, and $\delta J^{s}_{L\to R}(2;k)$ arises solely from the interacting 2-photon bound state  \cite{DRoy_nonlinearPRA2012,SM}. 

\begin{figure}[h]
    \centering
\includegraphics[width=0.48\textwidth]{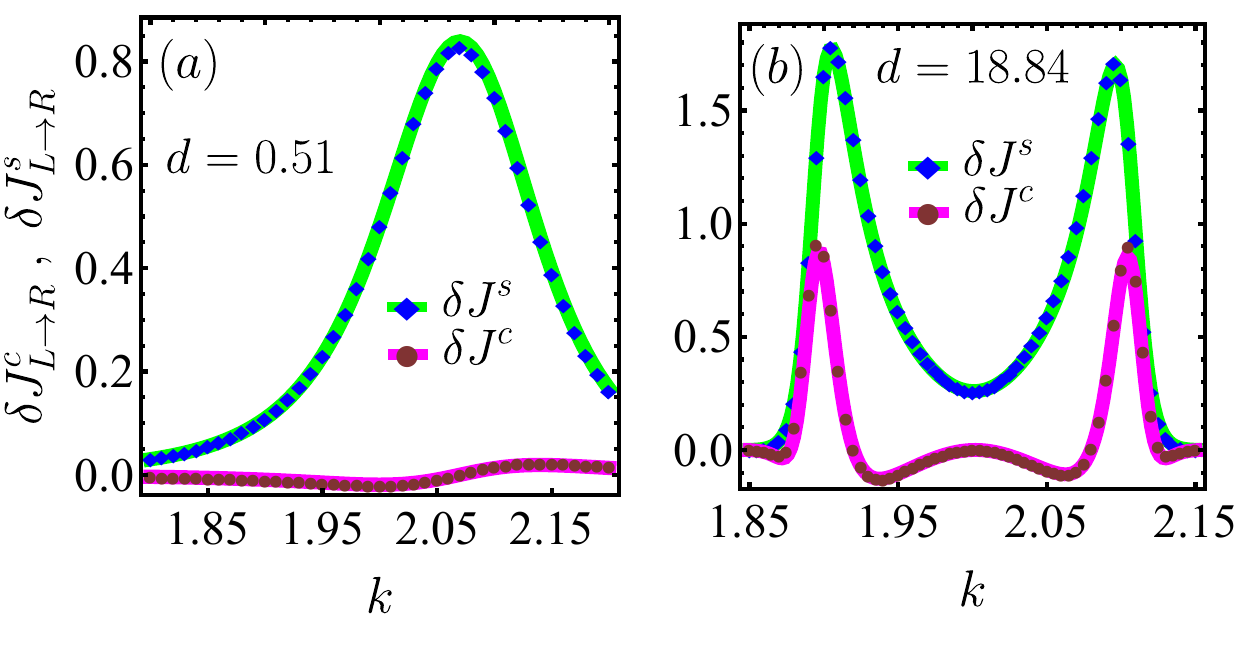}
    \caption{Nonlinear current components $\delta J^{c}_{L\to R}(2;k)$, and $\delta J^{s}_{L\to R}(2;k)$ as functions of incident photons' momentum $k$ for two different separation $(d)$ values. Cross-correlation current contributes significantly in the non-Markov regime (e.g.,  $d=18.84$) of a giant atom. Solid lines represent numerical integration, and dotted points are obtained by analytical series sums including contributions from the roots $\{p_j,q_j\}$ within the window $\Re[p_j],\Re[q_j]\in [\Omega-15\Gamma_{k=0},\Omega+15\Gamma_{k=0}]$. We set $g_1=g_2=0.2,$ and  $\Omega=2.0$.}
\label{Fig3}
\end{figure}

The cross-correlation part of the 2-photon transmission current is given as follows:
\begin{align}
   \delta J^{c}_{L\to R}(2;k)
   = \frac{2}{\pi}\Im\bigg[ \frac{(t^R_{k})^{*} \Gamma_{k}|\phi^{R}_b(k)|^2}{f^{2}_{+}(k) \langle b;b|\hat{G}_{o}^{+}(2k)|b;b\rangle }\bigg], \label{Eq_GiantCross}
\end{align}
$\Im [...]~(\Re [...])$ denotes the imaginary (real) part of the expression. We take two limits of the above expression. In the small atom limit, when $g_1=0$ or $g_2=0$, we get $\delta J^{c}_{L\to R}(2;k)=0.$ Furthermore, in the Markov regime of the giant atom, we replace 
the momentum-dependent phase $kd$ in all 2-photon integrations by a constant phase $\Omega d$, and arrive at $\delta J^{c}_{L\to R}(2;k)\approx 0.$ A significant deviation of $\delta J^{c}_{L\to R}(2;k)$ from zero arises due to non-Markov effect for two incident photons with same momentum,  as shown in Fig.~\ref{Fig3}(a-b). After a slight simplification of the terms inside the brackets in Eq.~\ref{Eq_GiantCross}, we find that the nonzero contributions originate from $\Im[ f_{+}(k)\langle b;b|\hat{G}_{o}^{+}(2k)|b;b\rangle ]^{-1}$, which serves as a measure of deviation of the cross-correlation current from the Markov regime. 

The second term $\delta J^{s}_{L\to R}(2;k)$ of the nonlinear transmission current originates exclusively from the 2-photon bound-state contribution and is given by 
\begin{align}
    \delta J^{s}_{L\to R}(2;k)
    =\frac{|\phi^{R}_b(k)|^4}{|\langle b;b|\hat{G}_{o}^{+}(2k)|b;b\rangle|^2} \mathcal{T}(2k),
\end{align}
where the integral $ \mathcal{T}(2k)=32\pi\int dk_1 ~|\phi^{R}_b(k_1)\phi^{R}_b(2k-k_1)|^2$
encodes atomic correlations that depend only on the modulus of $\phi^{R}_b(k)$.
We can express $\mathcal{T}(2k)$ in terms of a series using the  roots $\{p_j\}$ and  $\{q_j=2k-\Re[p_j]+\Im[p_j]\}$  in \cite{SM}.
The {\it exact} 2-photon current $ J_{L\to R}(2;k)$  is invariant under the transformation $g_1 \leftrightarrow g_2$. Therefore, 
regardless of its inherent structural asymmetry (when $g_1\neq g_2$), the giant atom demonstrates robust nonlinear reciprocal transport. This is a surprising result when we compare it with nonreciprocal nonlinear light transmission in a side-coupled WQED setup of two spatially separated atoms with broken structural symmetry \cite{Hamann2018}.  In the symmetric configuration of this two-atom WQED setup, the moduli of the 1-photon excitation amplitudes are related by $|\phi^{R}_{b_1}(k)|= |\phi^{L}_{b_2}(k)|$ and $|\phi^{R}_{b_2}(k)|= |\phi^{L}_{b_1}(k)|$. Here, $b_1$ and $b_2$ are atomic labels.
For a single giant atom, we find that  $|\phi^{R}_{b}(k)|= |\phi^{L}_{b}(k)|$ always holds due to 1-photon interference effects, and this condition is sufficient to yield a reciprocal nonlinear transport in this setup.
In the Markov regime of the giant atom, 
 we recover the following expression;
\begin{align}
     \delta J^{s}_{L\to R}(2;k)\approx \frac{\Gamma^{3}_{\Omega}}{2(2\pi)^2 [(k-\Omega-\Delta_{\Omega})^2+\Gamma^{2}_{\Omega}/4]^2},
\end{align}
which remains invariant under $g_1 \leftrightarrow g_2$, again showing a reciprocal 2-photon transport. 
Next, we propose a Markov WQED setup where we regulate the interference effect and demonstrate control over nonreciprocal to reciprocal transport regimes in a generalized direct-coupled configuration.

\begin{figure}[h]
    \centering \includegraphics[width=8.6cm, height=2.3cm]{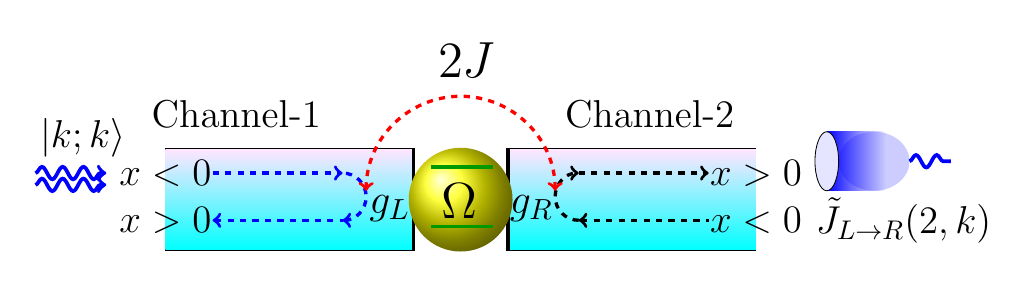}
    \caption{A generalized direct-coupled atom-waveguide setup with interference effects  regulated via tunneling $2J$.  Two photons in the product state, with momentum $k$, are being injected into channel-1. A photo-detector measures the transmitted 2-photon current on the other side.  Here, $|g_L|\neq |g_R|$ induces a structural asymmetry in the system. }
    \label{Fig_DirectCoupled}
\end{figure}

{2. \bf Generalized direct-coupled atom-waveguide setup:} In the generalized direct-coupled WQED setup, the atom in the middle is connected to two semi-infinite waveguides at both ends, which are, respectively, channel-$1$ and channel-$2$ on the left and right ends \cite{DRoy_OpticalDiode_PRB2010, RMP_DRoy2017}. The incoming and outgoing paths in the two channels are denoted by the coordinates $x<0$ and $x>0$, respectively (see Fig.~\ref{Fig_DirectCoupled}). 
We denote $\hat{a}^{\dagger}_{i}(x)$ as the photon creation operator in the $i~(=1,2)$th channel at location $x$. 
The Hamiltonian for the full setup can be written as 
\begin{align}   
    \hat{H}^d =\hat{H}_b&+\hat{H}'_{\text{w}}+\Big[\big(g_L \hat{a}^{\dagger}_{1}(0)+g_R \hat{a}^{\dagger}_{2}(0)\big)\hat{b}+\text{H.c.}\Big]\nonumber\\
    &+2J \hat{a}^\dagger_{1}(0)\hat{a}_{2}(0)+2J^*\hat{a}^\dagger_{2}(0)\hat{a}_{1}(0),\label{Hamiltonian_dcouple}
\end{align}
where the Hamiltonian for the waveguides is given by
\begin{align}
    \hat{H}'_{\text{w}}=-i&\int_{-\infty}^{\infty} dx \Big[\hat{a}^{\dagger}_{1}(x)\partial_{x} \hat{a}_{1}(x)+\hat{a}^{\dagger}_{2}(x)\partial_{x} \hat{a}_{2}(x)\Big].
\end{align}
We incorporate an additional tunneling $2J$, which controls the direct coupling between channel-$1$ and channel-$2$,  without involving the atom. This  mimics a regulated interference path in the setup without a time lag. The atom is coupled asymmetrically to the left and right waveguides with amplitudes $g_L$ and $g_R$, when $|g_L|\neq |g_R|$. Here, we consider $g_L,g_R$ and $J$ to be complex.

{\it 1-photon paradigm.}-- The excitation amplitude of the atom by a single photon with momentum $k$ incident from the left is given by:
\begin{gather}
  \tilde{\phi}_{b}^{R}(k)=\frac{1}{\sqrt{2\pi}}\frac{  ( g_L^*-iJ^*g_R^*)/(|J|^2+1)}{({k}-\Omega-{\Delta})+i{\Gamma}/2}, \label{Eq_exciationDirect}
\end{gather}
where the Lamb shift $({\Delta})$ and the relaxation rate $({\Gamma})$ of the atom are given by ${\Delta}=-(g_L g_R^* J^*+g_L^* g_R J)/(2|J|^2+2)$ and ${\Gamma}=(|g_L|^2+|g_R|^2)/(|J|^2+1)$, respectively.
We observe that the Lamb shift for such a small atom in the linear waveguides vanishes whenever we disconnect the extra interference path by setting $J=0$ \cite{PhysRevA_Bag2023}. The self-energy terms $(\Delta, \Gamma)$ are invariant under the transformation: $g_L \leftrightarrow g_R$, $J\to J^*$. This transformation maps quantities associated with right-moving incident photons (labeled $R$) to those of left-moving incident photons (denoted $L$). Using Eq.~\ref{Eq_exciationDirect}, we find that the modulus of the excitation amplitude $\tilde{\phi}_{b}^{R}(k)$ can differ under such a transformation, 
leading to $|\tilde{\phi}_{b}^{R}(k)| \neq |\tilde{\phi}_{b}^{L}(k)|$ for the generic parameters. The transmission amplitude of a single photon reads as follows:
\begin{align}
\tilde{t}^{R}_{k}=-i\frac{  [ 2J^* ({k}-\Omega)+ g_Rg_L^* ]/(|J|^2+1)}{({k}-\Omega-{\Delta})+i{\Gamma}/2},
\end{align}
which becomes $\tilde{t}^{R}_k= \tilde{t}^{L}_k$ only when $J$, $g_L$ and $g_R$ are real. However, in all regimes, $|\tilde{t}^{R}_k|^2=|\tilde{t}^{L}_k|^2$ indicates a reciprocal transport of one photon in a two-port system. In the absence of atom-photon coupling, i.e., when $g_L=g_R=0$, we get $\tilde{t}^{R}_{k}=-2iJ^{*}/(|J|^2+1)$. Consequently, the values $J=\pm 1, \pm i$ play a  special role, allowing for perfect transmission of single photons from the left to right waveguide with zero atom-photon coupling. 

{\it 2-photon paradigm.}-- We obtain various Green's functions associated with the interacting 2-photon bound state in \cite{SM}.  Once again, the transmission current of two non-interacting photons from the left to right waveguide is given by $\tilde{J}^{o}_{L\to R}(2;k)=(|\tilde{t}^R_{k}|^2/\pi) \delta(0),$ where the 
1-photon current is given by $|\tilde{t}^R_{k}|^2/2\pi$. As before, the transmission current of two interacting photons has two parts defined by $\delta \tilde{J}_{L\to R}(2;k)=\delta \tilde{J}^{c}_{L\to R}(2;k)+\delta \tilde{J}^{s}_{L\to R}(2;k)$. The cross-correlation part is
\begin{align}
    &\delta \tilde{J}^{c}_{L\to R}(2;k)\nonumber\\
    =&\frac{4}{\pi} \Im\Bigg[ \frac{\tilde{t}^{R*}_{k}  (g_R-i  J^* g_L)(g^*_L-i  J^* g^{*}_R)}{(k-\Omega-{\Delta}+i{\Gamma}/2)(|J|^2+1)^2}  |\tilde{\phi}_{b}^{R}(k)|^2\Bigg].
\end{align}
For generic parameters with arbitrary complex coupling amplitudes, we find that the cross-correlation current is nonzero, and $\delta \tilde{J}^{c}_{L\to R}(2;k)\neq \delta \tilde{J}^{c}_{R\to L}(2;k)$. 
The 2-photon current due to an interacting 2-photon bound state is given by
\begin{align}
    \delta \tilde{J}^{s}_{L\to R}(2;k)=\frac{4|g_R-i J^{*} g_L|^2}{(|J|^2+1)^2}|\tilde{\phi}^{R}_{b}(k)|^4,
\end{align} 
which once again gives  
$\delta \tilde{J}^{s}_{L\to R}(2;k)\neq \delta \tilde{J}^{s}_{R\to L}(2;k)$ for generic parameters where $|\tilde{\phi}_{b}^{R}(k)| \neq |\tilde{\phi}_{b}^{L}(k)|$. Therefore, an asymmetric excitation probability of the  atom in the generalized direct-coupled setup gives a nonreciprocal nonlinear transport. However, nontrivial parameter regimes can be found, where the reciprocity emerges in 2-photon transport. 
\begin{figure}[h]
    \centering
\includegraphics[width=8.65
cm, height=3.3cm]{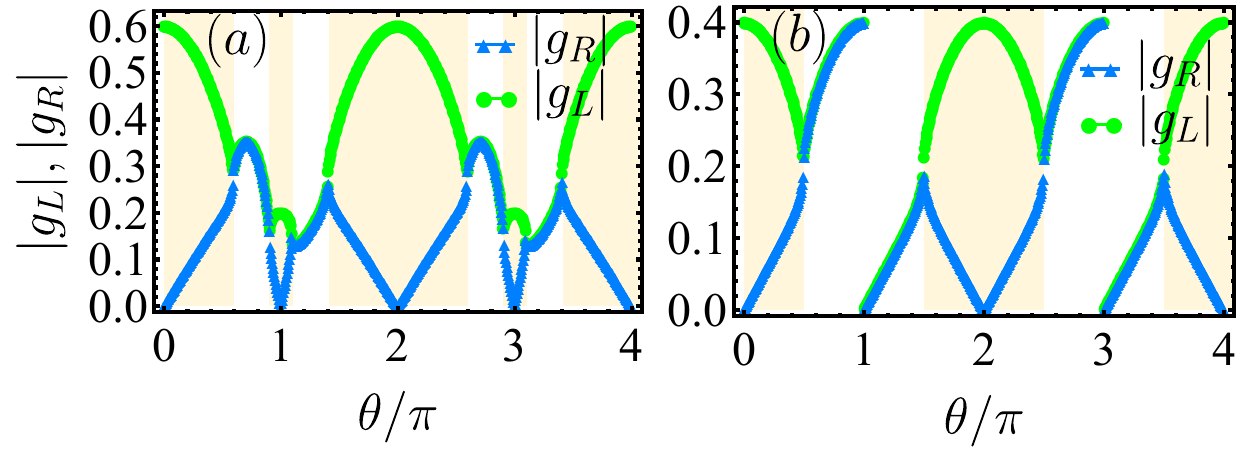}
    \caption{ We plot the absolute value of $g_L,$ and $g_R$ as function of the free parameter $\theta$ from the mapping relations in Eqs.~(\ref{Eq_Transformation_A}-\ref{Eq_Transformation_B}). The highlighted sections show the nontrivial regions where $|g_L|\neq |g_R|$ by the mapping relations. We set $g_1=0.2, g_2=0.1$ in (a), and $g_1=g_2=0.1$ in (b). }
\label{Fig4}
\end{figure}

{Reciprocal transport regimes.--}   We examine two cases for the non-trivial parameter regimes in which reciprocal transport is revived in the generalized direct-coupled setup.
Case (a). We find that the simplest setting in which 2-photon transport can be tuned to the reciprocal regime is with $J=\pm 1$, and keeping $g_L \neq g_R$ for the real values of $g_L$ and $g_R$.  In this case, we get $|\tilde{\phi}_{b}^{R}(k)|=|\tilde{\phi}_{b}^{L}(k)|$ in a parity-broken configuration, along with $\tilde{t}^{R}_k=\tilde{t}^{L}_k$. 

Case (b). If we consider complex parameters with nonzero imaginary values, then reciprocal 2-photon transport emerges by the following choice of the coupling parameters: 
\begin{align}
     &g_L=\frac{e^{i {\theta}/2}}{g_1+g_2e^{i {\theta}}}\Big[g_1^2+g_2^2+2g_1g_2e^{i{\theta}}
     + h(g_1,g_2,\theta)\Big],\label{Eq_Transformation_A}\\
     &g_R=\frac{e^{i {\theta}/2}}{g_1+g_2e^{i {\theta}}}\Big[g_1^2+g_2^2+2g_1g_2e^{-i\theta}- h(g_1,g_2,\theta)\Big]\label{Eq_Transformation_B},
\end{align}
along with $J=-i$. Here, $g_1,g_2$ are the real coupling amplitudes in the corresponding giant atom-waveguide setup.  
In the above equations, we use 
\begin{align}
h(g_1,g_2,\theta)=\sqrt{(\Gamma_{\theta}/2)^2-4g_1^2g_2^2\sin^2{\theta}}.\label{Eq_Transformation_C}
\end{align}
Within these parameter regimes, we get $|\tilde{\phi}_{b}^{R}(k)|=|\tilde{\phi}_{b}^{L}(k)|,$ and $\tilde{t}^{R}_k=-\tilde{t}^{L}_k$.
Now, the setup can still  maintain a structural asymmetry with $|g_L|\neq |g_R|$ for some values of the real phase $\theta$, as shown in
Fig.~\ref{Fig4} (a-b).
The highlighted (colored) sections in the plot signify the nontrivial regions where $|g_L|\neq |g_R|$.  We notice that whenever $|g_L| \neq |g_R|$, the revival of reciprocity requires perfect tunneling of photons $(|J|=1)$ between the two semi-infinite waveguides via the regulated interference path.
We use  $\theta=k$ in Eqs.~\ref{Eq_Transformation_A}–\ref{Eq_Transformation_C} to facilitate a mapping between the side-coupled giant atom setup and the generalized direct-coupled setup in the 1-photon regime \cite{PhysRevA_sideCoupledDirectcoupled,PhysRevA_Bag2023}. In this setting, 1-photon transmission amplitudes, for right-moving photons, are matched exactly in the two setups, and so are the atomic excitation amplitudes : $t^{R}_k=\tilde{t}^{R}_k$, and $\phi^{R}_k=\tilde{\phi}^{R}_k$. This further gives identical 1-photon and (linear) 2-photon  transmission currents in the two setups.  However, 
the mapping cannot capture all regimes of the total inelastic scattering involving two photons in the non-Markov giant atom-waveguide system. So, the {\it exact} nonlinear 2-photon currents in these setups are not mapped onto each other. Interestingly, we can recover the approximate expressions for the nonlinear current in the Markov regimes of the giant atom using  $\theta=\Omega$ in the mapping Eqs.~\ref{Eq_Transformation_A}–\ref{Eq_Transformation_C} (see \cite{SM}).

Therefore, we witness a revival of reciprocity in this generalized nonlinear direct-coupled setup. Here, we have different occupation statistics for excitations in the atom than in the photonic waveguides, and parametric structural asymmetry. However, the setup fails to yield nonreciprocal transport when the 1-photon excitation amplitudes satisfy $|\tilde{\phi}_{b}^{R}(k)| = |\tilde{\phi}_{b}^{L}(k)|$ due to interference effects. 

{\it Scattering matrix.}-- We now analyze the elastic and inelastic components in the  
scattering $\mathbf{S}$-matrix to see the signatures of a nonreciprocal-to-reciprocal transmission transition in the generalized direct-coupled setup. The  $\mathbf{S}$-matrix maps an incoming plane-wave state to an outgoing plane-wave state. The 2-photon plane-wave states are denoted by $|S_{k_1,k_2}(l,l')\rangle=\int \int dx_1 dx_2 \frac{e^{i(k_1x_1+k_2x_2)}}{2\pi }\hat{a}^\dagger_{l}(x_1) \hat{a}^\dagger_{l'}(x_2)|\varphi\rangle$ for channels $l,l'=1,2$. The mapping generated by the $\mathbf{S}$-matrix is given by the following expression \cite{ShenFanPRL2007}:
\begin{align}
    &\langle S_{q_1,q_2}(m,m')|\hat{\mathbf{S}}|S_{k_1,k_2}(l,l')\rangle\nonumber\\   =&\Big[\mathbf{M}_{m,m'}^{l,l'}(k_1,k_2)\delta(q_1-k_1) \delta(q_2-k_2)+(k_1 \leftrightarrow k_2)\Big]\nonumber\\
    &+ \mathbf{N}_{m,m'}^{l,l'}({\mathbf q};{\mathbf k})\delta(E_{\mathbf q}-E_{\mathbf k}).
\end{align}
We use $\mathbf k=(k_1,k_2)$ and $\mathbf q=(q_1,q_2)$ for the incident and scattered momenta of two photons.  Here, $E_{\mathbf k}=k_1+k_2$ denotes the total energy of two incident photons (similarly for $E_{\mathbf q}$). 
The first two terms (direct and exchange) in the above expression describe the momentum distribution of the two photons involved in elastic scattering. 
The last term, on the other hand, yields inelastic scattering processes. 
We arrange elements $\mathbf{M}_{m,m'}^{l,l'}(k_1,k_2)$ and $\mathbf{N}_{m,m'}^{l,l'}({\mathbf q};{\mathbf k})$ in the form of $3\times 3$ matrices denoted by $\mathbf{M}(k_1,k_2)$ and $\mathbf{N}({\mathbf q};{\mathbf k})$, where $(m,m')$ and $(l,l')$ run over the set $((1,1), (1,2), (2,2))$. The subscripts $(m, m')$ denote the outgoing 2-photon channels and the superscripts $(l,l')$ indicate the incoming 2-photon channels. 
 We have 
\begin{align}
  \mathbf{M}(k_1,k_2)= 
  \begin{bmatrix}
        \tilde{r}^R_{k_1}  \tilde{r}^R_{k_2}  &  \tilde{r}^R_{k_1}  \tilde{t}^L_{k_2}& \tilde{t}^L_{k_1}  \tilde{t}^L_{k_2} \\
         \tilde{t}^R_{k_1}  \tilde{r}^R_{k_2} & \tilde{t}^R_{k_1}  \tilde{t}^L_{k_2} & \tilde{r}^L_{k_1}  \tilde{t}^L_{k_2} \\
         \tilde{t}^R_{k_1}  \tilde{t}^R_{k_2} &  \tilde{t}^R_{k_1}  \tilde{r}^L_{k_2} &\tilde{r}^L_{k_1}  \tilde{r}^L_{k_2}
    \end{bmatrix}, \label{M_matrix_paper_2photon}
\end{align}
where  $\tilde{r}^R_{k}$ and $\tilde{r}^L_{k}$ are the reflection amplitudes of a single right- and left-moving photon with momentum $k$ (see \cite{SM}). We observe  $|\mathbf{M}_{2,2}^{1,1}(k_1,k_2)|^2=|\mathbf{M}_{1,1}^{2,2}(k_1,k_2)|^2$ and $|\mathbf{M}_{1,2}^{1,1}(k_1,k_2)|^2=|\mathbf{M}_{1,2}^{2,2}(k_1,k_2)|^2$; which signal that the probability distribution for the elastically scattered photons across different channels remains the same for light injected from either the left or right side of the atom. The elements $\mathbf{N}_{m,m'}^{l,l'}({\mathbf q};{\mathbf k})$ can be written in matrix form as follows:
\begin{widetext}
{
\begin{align}
  &\mathbf{N}({\mathbf q};{\mathbf k})= \begin{bmatrix}
        (g_L-i J g_R)^2(g^*_L-i J^*g^*_R)^2  &  (g_L-i J g_R)^2(g^*_L-i J^*g^*_R)(g^*_R-i J g^*_L)& (g_L-i J g_R)^2(g^*_R-i J g^*_L)^2 \\
        &&\\
         \splitfrac{(g_L-i J g_R) (g_R-i J^*g_L)}{\times(g^*_L-i J^*g^*_R)^2} & \splitfrac{(g_L-i J g_R) (g_R-i J^*g_L)}{\times(g^*_L-i J^*g^*_R)(g^*_R-i J g^*_L)} & \splitfrac{(g_L-i J g_R) (g_R-i J^*g_L)}{\times(g^*_R-i J g^*_L)^2} \\
         &&\\
         (g_R-i J^*g_L)^2(g^*_L-i J^*g^*_R)^2 &  (g_R-i J^*g_L)^2(g^*_L-i J^*g^*_R)(g^*_R-i J g^*_L) & (g_R-i J^*g_L)^2(g^*_R-i J g^*_L)^2
    \end{bmatrix}h({\mathbf q};{\mathbf k}),\label{N_matrix_paper_2photon}
\end{align}}
\end{widetext}
where $h(\mathbf q;\mathbf k)=-\frac{16i}{\pi} \frac{A/(|J|^2+1)^4}{[(k_1-k_2)^2-A^2][(q_1-q_2)^2-A^2]}$ and $A=(E_{\mathbf k}-2\Omega-2\Delta+i\Gamma)$.
We observe that the following two inequalities hold for generic parameters: $|\mathbf{N}_{2,2}^{1,1}({\mathbf q};{\mathbf k})|^2\neq|\mathbf{N}_{1,1}^{2,2}( {\mathbf q}; {\mathbf k})|^2,$ and $ |\mathbf{N}_{1,2}^{1,1}( {\mathbf q}; {\mathbf k})|^2 \neq|\mathbf{N}_{1,2}^{2,2}( {\mathbf q}; {\mathbf k})|^2$ when the light transmission is nonreciprocal and $|\tilde{\phi}_{b}^{R}(k)| \neq |\tilde{\phi}_{b}^{L}(k)|$. In such regimes, the probability distribution of two inelastically scattered photons across different channels depends on whether the photons are incident from the left or right side of the atom \cite{DRoy_OpticalDiode_PRB2010}. 
We recall the special parameter regimes: (a) when all the parameters $J,g_L$ and $g_R$ are real (keeping $g_L \neq g_R$), along with  $J= \pm 1$; and (b) when the parameters are complex and satisfy Eqs.~\ref{Eq_Transformation_A}–\ref{Eq_Transformation_C} with $J=-i$. In both cases (a) and (b), we obtain $|\mathbf{N}_{2,2}^{1,1}({\mathbf q};{\mathbf k})|^2=|\mathbf{N}_{1,1}^{2,2}( {\mathbf q}; {\mathbf k})|^2$ and $|\mathbf{N}_{1,2}^{1,1}( {\mathbf q}; {\mathbf k})|^2 =|\mathbf{N}_{1,2}^{2,2}( {\mathbf q}; {\mathbf k})|^2$, i.e., the distribution of all scattered photons are independent of the direction of incoming photons, and hence 2-photon transmission becomes reciprocal.


 {\bf Conclusion:}  In summary, we have investigated a mechanism for reciprocal light transmission in structurally asymmetric nonlinear devices via 1-photon interference. \textcite{Fan_OpticalIsolatorlimitations_2015Nat} have demonstrated dynamic reciprocity for counter-propagating waves with non-overlapping frequencies in classical nonlinear optical isolators made of asymmetric structures. In contrast, we propose a mechanism based on a fully quantum treatment in which 1-photon interferences arising from multiple propagation pathways within the system play a central role. We find that, due to interfering pathways, a single right- or left-incident photon can give the same excitation probability of the two-level atom, and it is enough to suppress the directional response of the excited atom in 2-photon processes in these studied models. 
Although our analysis is carried out for a 2-photon input state, the findings can be extended to weak coherent-state inputs \cite{PhysRevA_singlevsCoh_Atul2023}, which are more common in WQED experiments. 



\bibliography{references}

\onecolumngrid
\vspace{25cm}
\begin{center}
{\large \bf{Supplementary Material for ``Revival of transport reciprocity via quantum interference
in asymmetric nonlinear devices"}} \\
\vspace{2mm}
{Rupak Bag and Dibyendu Roy} \\
{Raman Research Institute, Bangalore 560080, India}
\end{center}
\vspace{0.1cm}

\onecolumngrid
\setcounter{figure}{0}
\setcounter{equation}{0}
\setcounter{page}{1}

\section{Giant atom-waveguide setup}

In the manuscript, we discuss a giant atom-waveguide setup. In this setup, an artificial two-level atom is side-coupled to an infinitely long waveguide at points $x=d/2$ and $x=-d/2$. The coupling amplitudes at these two points are $g_1$ and $g_2$, respectively. We take  $g_1$ and $g_2$ to be reals. As discussed in the manuscript, we represent the two-level atom using an anharmonic site-b with Kerr nonlinearity $U$. We take $U\to \infty$ to mimic the two-level behavior. Hamiltonian for the whole system  is ($ \hbar=1$)
\begin{align}   &\hat{H}^{g}=\hat{H}_{\text{b}}+\hat{H}_{\text{w}}+ \int_{-\infty}^{\infty} dx~ g(x)\Big[\big(\hat{a}^\dagger_{R}(x)+\hat{a}^\dagger_{L}(x)\big)\hat{b}+\text{H.c.}\Big], \label{Eq_Hamiltonian_giantatom_Supple}\\
   &\hat{H}_b=\Omega\hat{b}^\dagger \hat{b}+\frac{U}{2}\hat{b}^\dagger \hat{b}^\dagger \hat{b}\hat{b},\\ &\hat{H}_{\text{w}}=-i v_g\int_{-\infty}^{\infty} dx \big[\hat{a}_R^\dagger(x)\partial_x \hat{a}_R(x)-\hat{a}_L^\dagger(x)\partial_x \hat{a}_L(x)\big],
\end{align}
where,  $g(x)=g_1\delta(x-d/2)+g_2 \delta(x+d/2)$. In the following subsections, we derive the transmission photon current operator and study light transport using 1-photon and  2-photon states in the waveguide. 

\subsection{Transmission  current operator}
In the manuscript, we use the transmission current operator to calculate the average 1-photon and 2-photon currents transmitted across the giant atom when it is asymmetrically coupled to the waveguide $(g_1\neq g_2)$. Here, we derive the expression for the transmission current operator for photons using the continuity equation. We can  define local density for the photons inside the waveguide using $\hat n(x)= \hat{a}^\dagger_{R}(x)\hat{a}_{R}(x)+ \hat{a}^\dagger_{L}(x)\hat{a}_{L}(x)$ at location $x$. 
We utilize the Heisenberg equations of motion for the field operators, such as
\begin{align}
    \frac{\partial}{\partial t}  \hat{a}_{R}(x)&=i[\hat{H}^{g},\hat{a}_{R}(x)] =v_g\int_{-\infty}^{\infty} dx' \Big[\hat{a}_R^\dagger(x'),\hat{a}_{R}(x)\Big]\partial_{x'} \hat{a}_R(x')+v_g\int_{-\infty}^{\infty} dx'~\hat{a}_R^\dagger(x')\Big[\partial_{x'} \hat{a}_R(x'),\hat{a}_{R}(x)\Big]-i g(x) \hat{b}.
\end{align}
After simplification, we get:
\begin{align}
    \frac{\partial}{\partial t}  \hat{a}_{R}(x)=-v_g\partial_{x} \hat{a}_R(x)-i g(x) \hat{b}, ~~\text{and similarly,}~~\frac{\partial}{\partial t}  \hat{a}_{L}(x)=v_g\partial_{x} \hat{a}_L(x)-i g(x) \hat{b}.
\end{align}
The rate of change of local photon density with time $t$ is governed by the following continuity equation 
\begin{align}
    \frac{\partial }{\partial t}\hat n(x)
    &=-v_g\partial_x \Big[\hat{a}^\dagger_{R}(x)\hat{a}_{R}(x)-\hat{a}^\dagger_{L}(x)\hat{a}_{L}(x)\Big]+ig(x)\Big[\hat{b}^\dagger \Big(\hat{a}_{R}(x)+\hat{a}_{L}(x)\Big)-\Big(\hat{a}^\dagger_{R}(x)+\hat{a}^\dagger_{L}(x)\Big) \hat{b}\Big].
\end{align}
For $x> d/2$, we have $g(x)=0$. Therefore, we rewrite the continuity equation in the form of $ \partial_{t}\hat n(x)=-\partial_x \hat{J}_{L\to R}(x),$
where the transmission current operator $\hat{J}_{L\to R}(x)$ is defined by
\begin{align}
    \hat{J}_{L\to R}(x)=v_g\Big[\hat{a}^\dagger_{R}(x)\hat{a}_{R}(x)-\hat{a}^\dagger_{L}(x)\hat{a}_{L}(x)\Big],\label{S_GiantAtom_currentoperator}
\end{align} 
where we consider that the photons are incident from the left of the atom and $x>d/2$. We choose $v_g=1$ in the following treatment and measure all energies in the unit of $v_g$.
\subsection{1-photon paradigm}
To study 1-photon transport, we write a scattering state of $\hat{H}^{g}$ in the single-excitation sector as follows:
\begin{align}
|\Phi_{1}(\alpha,k)\rangle=&\int_{-\infty}^{\infty} dx \bigg[\phi_R^\alpha(x,k) \hat{a}^\dagger_R(x)+\phi_L^\alpha(x,k)\hat{a}^\dagger_L(x)\bigg]|\varphi;0\rangle+\phi_{b}^{\alpha}(k) \hat{b}^\dagger|\varphi;0\rangle.
\end{align}
Here, $\alpha =R,L$ denotes  the initial condition where photons are propagating either in the right ($R$)- or left ($L$)-moving channel. $|\varphi;0\rangle$ indicates the vacuum state of the 
fields inside the waveguide with atom in the ground state. The momentum of the incident photon is $k$. From the time-independent Schr\"{o}dinger equation, $\hat{H}^{g}|\Phi_{1}(\alpha,k)\rangle={k}|\Phi_{1}(\alpha,k)\rangle$, we  obtain:
\begin{align}
  & \big(-i\partial_x-{k}\big)\phi_{R}^{\alpha}(x,k)+g(x) \phi_{b}^{\alpha}(k)=0, \label{G_1}\\
  & \big(i\partial_x-{k}\big)\phi_{L}^{\alpha}(x,k)+g(x) \phi_{b}^{\alpha}(k)=0, \label{G_2}\\
   & \big(\Omega-{k}\big)\phi_{b}^{\alpha}(k)+\int_{-\infty}^{\infty} dx~g(x) \Big[\phi_{R}^{\alpha}(x,k)+\phi_{L}^{\alpha}(x,k)\Big]=0. \label{G_3}
\end{align}
 To account for the discontinuous jumps in the wavefunctions across the coupling points $x=\pm d/2$, we use 
\begin{align}
    \phi_{R}^{\alpha}\Big(\pm\frac{d}{2},k \Big)=\frac{1}{2}\bigg[\phi_{R}^{\alpha}\Big(\pm\frac{d}{2}^{+},k \Big)+\phi_{R}^{\alpha}\Big(k,\pm\frac{d}{2}^{-} \Big)\Bigg],~\text{and}~\phi_{L}^{\alpha}\Big(\pm\frac{d}{2},k \Big)=\frac{1}{2}\bigg[\phi_{L}^{\alpha}\Big(\pm\frac{d}{2}^{+},k \Big)+\phi_{L}^{\alpha}\Big(k,\pm\frac{d}{2}^{-} \Big)\Bigg].
\end{align}

{ \textbf{\textrm{I}. Incident photon in $R$-channel}.--}
We consider a single photon incident from the left of the atom. The photon is propagating with momentum $k$ in the  $R$-channel. The ansatz for  the scattered wavefunction amplitudes can be written as follows:
\begin{align} 
\phi_R^{R}(x,k)=&\frac{e^{ikx}}{\sqrt{2\pi}}\big[\Theta(-x-d/2)+u^{R}_k\Theta(x+d/2)\Theta(-x+d/2)+ t^{R}_k\Theta(x-d/2)\big],\nonumber\\
\phi_L^{R}(x,k)=&\frac{e^{-ikx}}{\sqrt{2\pi}}\big[r^{R}_k\Theta(-x-d/2) + w^{R}_k\Theta(x+d/2)\Theta(-x+d/2)\big],
\end{align}
In the above expressions, $\Theta(x)$ denotes the Heaviside step function. Using the ansatz in Eq.~\ref{G_1}-\ref{G_3}, we obtain 
\begin{gather}
    t^R_k=\frac{({k}-\Omega-\Delta_{k})}{({k}-\Omega-\Delta_k)+i\Gamma_k/2},
~~~~ r^R_k=-\frac{i(g_1e^{-ikd/2}+g_2e^{ikd/2})^2}{({k}-\Omega-\Delta_k)+i\Gamma_k/2},\label{single photon reflection and transmission amplitude}\\
    w^{R}_k=\frac{-i(g_1g_2+g_2^2e^{ikd})}{({k}-\Omega-\Delta_k)+i\Gamma_k/2},~~~
   u^{R}_k=\frac{{k}-\Omega+i(g_2^2+g_1g_2e^{ikd})}{({k}-\Omega-\Delta_k)+i\Gamma_k/2},~~~
   \phi_{b}^{R}(k)=\frac{1}{\sqrt{2\pi}}\frac{(g_1e^{-ikd/2}+g_2e^{ikd/2})}{({k}-\Omega-\Delta_k)+i\Gamma_k/2}.
\end{gather}
Here, $t^{R}_k$ and $r^{R}_k$ denote the transmission and reflection amplitudes of a single photon, and the atomic excitation amplitude is $\phi_{b}^{R}(k)$. The average 1-photon current transmitted across the atom is $J_{L\to R}(1;k)=\langle \Phi_{1}(R,k)|\hat{J}_{L\to R}(x)|\Phi_{1}(R,k)\rangle=|t^{R}_{k}|^2/2\pi.$
The momentum-dependent Lamb shift $(\Delta_k)$ and the relaxation rate $(\Gamma_k)$ of the atom appear in the denominator of the complex wavefunction amplitudes. They are given by
\begin{gather}
    \Delta_k=2g_1g_2\sin{kd},~\text{and}~
    \Gamma_k=2(g_1^2+ g_2^2+2g_1g_2\cos{kd}).
\end{gather}
We notice that the structural asymmetry $(g_1\neq g_2)$ affects $u^{R}_k,w^{R}_k$, and  the phase associated with the complex amplitude $\phi_{b}^{R}(k)$.

{ \textbf{\textrm{II}. Incident photon in $L$-channel}.--} When the photon is incident from the right side of the atom, we consider the following ansatz:
\begin{align} 
\phi_R^{L}(x,k)&=\frac{e^{ikx}}{\sqrt{2\pi}}\big[ w^{L}_k\Theta(x+d/2)\Theta(-x+d/2) + r^{L}_k\Theta(x-d/2)\big],\nonumber\\\
\phi_L^{L}(x,k)&=\frac{e^{-ikx}}{\sqrt{2\pi}}\big[ t^{L}_k\Theta(-x-d/2)+ u^{L}_k\Theta(x+d/2)\Theta(-x+d/2)+\Theta(x-d/2)\big].
\end{align}
The amplitudes are given by
\begin{gather}
t^L_k=\frac{({k}-\Omega-\Delta_{k})}{({k}-\Omega-\Delta_k)+i\Gamma_k/2},~~~~ r^L_k=-\frac{i(g_2e^{-ikd/2}+g_1e^{ikd/2})^2}{({k}-\Omega-\Delta_k)+i\Gamma_k/2},\\
    w^{L}_k=-\frac{i(g_1g_2+g_1^2e^{ikd})}{({k}-\Omega-\Delta_k)+i\Gamma_k/2},~~~u^{L}_k=\frac{{k}-\Omega+i(g_1^2+g_1g_2e^{ikd})}{({k}-\Omega-\Delta_k)+i\Gamma_k/2},~~~~\phi_{b}^{L}(k)=\frac{1}{\sqrt{2\pi}}\frac{(g_2e^{-ikd/2}+g_1e^{ikd/2})}{({k}-\Omega-\Delta_k)+i\Gamma_k/2}.
\end{gather}
The above treatment gives, $|t^R_k|^2=|t^L_k|^2$, i.e., reciprocal 1-photon transport. Furthermore, we get symmetric excitation probability of the atom, i.e.,  $|\phi_{b}^{R}(k)|^2=|\phi_{b}^{L}(k)|^2$ due to the interferences in the giant atom setup.

\subsection{2-photon paradigm}
In the manuscript, we use the Lippmann-Schwinger formalism to study transport properties of two correlated photons. Here, we present detailed calculations for that. We consider that the photons are incident from the left of the atom with momentum $k$.  The 2-photon  scattering state $|\Psi_{2}(R,k;R,{k})\rangle$  is given by the Lippmann-Schwinger equation:
\begin{align}
|\Psi_2(R,k;{R},{k})\rangle=&|\Phi_2(R,k;{R},{k})\rangle+\hat{G}^{+}_o(2k)\Big(\frac{U}{2}\hat{b}^\dagger\hat{b}^\dagger \hat{b} \hat{b}\Big)|\Psi_2(R,k;{R},{k})\rangle,\label{Lippmann_1g}
\end{align}
where $2k$ is the total energy of the incident photons. Here,  $\hat{G}^{+}_o(2k)=(2k-\hat{H}^{g}_o+i0^{+})^{-1}$ is the retarded Green's function operator of the non-interacting Hamiltonian $\hat{H}^{g}_o=\hat{H}^{g}-U\hat{b}^\dagger\hat{b}^\dagger \hat{b} \hat{b}/2$.   The non-interacting 2-photon state $|\Phi(R,k;{R},{k})\rangle$ is given by
\begin{align}   |\Phi_{2}(R,k;{R},{k})\rangle=\frac{1}{\sqrt{2}}|\Phi_{1}(R,k)\rangle  |\Phi_{1}(R,k)\rangle, \label{S21_supple}
\end{align}
where the incident photons are injected as a product state into the waveguide. We denote the 2-photon computational basis state at the impurity site-b by $|b;b\rangle=(\hat{b}^\dagger)^2|\phi,0\rangle/\sqrt{2}$. In the two-level limit, 
we take $U\to \infty$ and  get 
\begin{align}
&|\Psi_{2}(R,k;{R},{k})\rangle=|\Phi_{2}(R,k;{R},{k})\rangle-\hat{{G}}^{+}_o(2k)|b;b\rangle \frac{\langle b;b|\Phi_{2}(R,k;{R},{k})\rangle}{\langle b;b|\hat{{G}}^{+}_o(2k)|b;b\rangle}=|\Phi_{2}(R,k;{R},{k})\rangle+| S_{2}(R,k;{R},{k})\rangle.\label{Lippmann_2g}
\end{align}
In the above sum, we separate the 2-photon scattering state into non-interacting $|\Phi_{2}(R,k;{R},{k})\rangle$ and interacting bound-state $| S_{2}(R,k;{R},{k})\rangle$ parts.

 \subsubsection{Green's function $\langle b;b|\hat{G}^{+}_{o}(E)|b;b\rangle$}
\begin{figure}[h!]
    \centering
\includegraphics[width=0.6\textwidth]{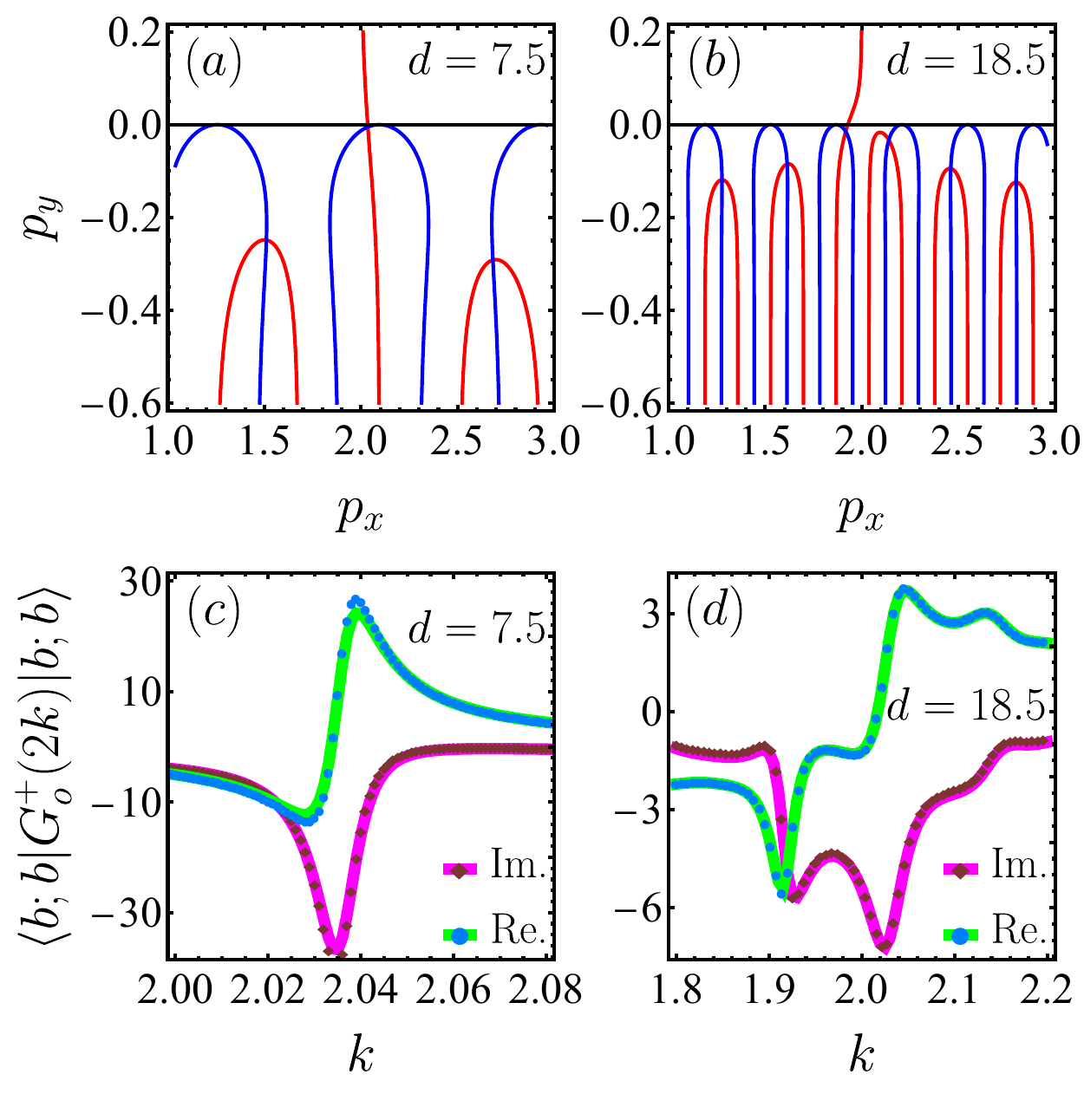}
    \caption{(a-b) Graphical representation of the  roots $\{p_j\}$ of the self-energy equation  at the intersection of the two contours: the red (blue) contour represents the real (imaginary) part of the complex equation $f_{+}(p)=0,$ with $p=p_x+ip_y$. The roots are unbounded (bounded) along $p_x$ ($p_y$). (c-d) Real and imaginary part of the Green's function $\langle b;b|\hat{G}_{o}^{+}(2k)|b;b\rangle$. The solid lines represent the numerical integration, and the dotted points are obtained from a restricted sum of the analytical series in Eq.~\ref{Eq_GreensFn_Analytical_supple}. In all plots, we choose $g_1=g_2=0.2$, and  $\Omega=2.0$. }
\label{Fig2}
\end{figure}

 In these calculations, we obtain an analytical series expression for the expectation of the 2-photon Green's function in the state $|b;b\rangle$. We show that $\langle b;b|\hat{G}^{+}_o(E)|b;b\rangle$ is symmetric in $g_1$ and $g_2$. 
 To this end,
we establish a 2-photon identity $\mathds{I}_2$ in the momentum space using  the non-interacting eigenstates of the  Hamiltonian $\hat{H}^{g}_o$ as follows:
\begin{align}
    \mathds{I}_2= \sum_{\alpha'_1,\alpha'_2=R,L}\int \int dk'_1 dk'_2~|\Phi_{1}(\alpha'_1,k'_1)\rangle  |\Phi_{1}(\alpha'_2,k'_2)\rangle\langle \Phi_{1}(\alpha'_1,k'_1)| \langle\Phi_{1}(\alpha'_2,k'_2)|.\label{I2_Giant}
\end{align}
Utilizing the 2-photon identity, we obtain
\begin{align}
    \langle b;b|\hat{{G}}^{+}_o(2k)|b;b\rangle 
    &=2\sum_{\alpha'_1,\alpha'_2}\int \int dk'_1 dk'_2 \frac{\big|\phi_{b}^{\alpha'_1}(k'_1)\phi_{b}^{\alpha'_2}(k'_2)\big|^2}{E-(k'_1+k'_2)+i 0^{+}}\nonumber\\
    &=\frac{2}{(2\pi)^2}\int    dk'_1 \frac{\Gamma_{k'_1}}{f_{+}(k'_1)f_{-}(k'_1)} \int dk'_2 \frac{1}{2k-(k'_1+k'_2)+i0^{+}} \frac{\Gamma_{k'_2}}{f_{+}(k'_2)f_{-}(k'_2)}. \label{EQ24_supple}
\end{align}
Let us first consider the $k'_2$ integration. We  use complex analysis and choose an integration contour that encloses the lower half of the complex plane. For this purpose, we define the two complex-valued functions: $f_{\pm}(p)=p-\Omega-\Delta_{p}\pm  i\Gamma_{p}/2=0$. The poles lying in the lower half of the complex plane are roots $\{p_j\}$ of the self-energy equation:  $f_{+}(p)=p-\Omega-\Delta_{p}+ i\Gamma_{p}/2=0$.
Now, the derivative function of $f_{+}(p)$ is
\begin{align}
    f'_{+}(p)=1-2g_1 g_2 d \big(\cos{p d}+i \sin{p d}\big)=1-2g_1 g_2 d e^{ip d}.
\end{align}
We have $f'_{+}(p=p_j) \neq 0$, since the additive quantity to unity is always a complex number as long as $\Im[p_j]\neq 0$. Therefore, the roots $\{p_j~|\forall j\in \mathbb{Z}^+\}$ are simple poles and contribute to the residues, giving   
\begin{align}
  \Rightarrow  \langle b;b|\hat{{G}}^{+}_o(2k)|b;b\rangle=& \frac{2(-2\pi i)}{(2\pi)^2}\int    dk'_1 \frac{\Gamma_{k'_1}}{f_{+}(k'_1)f_{-}(k'_1)}\sum_{j}\frac{1}{2k-(k'_1+p_j)+i0^{+}} \frac{\Gamma_{p_j}}{f'_{+}(p_j)f_{-}(p_j)}.
\end{align}
For the $k'_1$-integration, once again, we use an integration contour which encloses the lower half of the complex plane, and get
\begin{align}
  \Rightarrow  \langle b;b|\hat{{G}}^{+}_o(2k)|b;b\rangle=-\sum_{k,j}\frac{\Gamma_{p_k}}{f'_{+}(p_k)f_{-}(p_k)}\frac{2}{2k-(p_k+p_j)} \frac{\Gamma_{p_j}}{f'_{+}(p_j)f_{-}(p_j)}. \label{Eq_GreensFn_Analytical_supple}
\end{align}
The analytical summation in Eq.~\ref{Eq_GreensFn_Analytical_supple} is  carried with the roots $p_j, p_k$ (where $j,k\in \mathbb{Z}^+$) and Eq.~\ref{Eq_GreensFn_Analytical_supple} is symmetric under the transformation $g_1 \leftrightarrow g_2$. 
 Now, the giant atom setup lies  in the Markov regime when the distance $d \ll \lambda=2\pi v_g/\Omega$.  Within the Markov approximation, we replace the momentum-dependent phase $kd$ in all 2-photon integrations by a constant phase $\Omega d$ \cite{Wenju_PhysRevA2023Markov} and  get $\langle b;b|\hat{{G}}^{+}_o(2k)|b;b\rangle\approx (k-\Omega-\Delta_{\Omega}+i\Gamma_{\Omega}/2)^{-1}$.
In that case, only one root $p_{\text{Mar}}\approx\Omega+\Delta_{\Omega}-i\Gamma_{\Omega}/2$ significantly contributes to the sum in Eq.~\ref{Eq_GreensFn_Analytical_supple}. 
As we increase the separation $d$ between the two coupling points, we observe the roots $\{p_j\}$ become increasingly dense [see Fig.~\ref{Fig2}(a-b)] and contributions from   multiple roots become significant in the non-Markov regime.
 However, the sum  is highly convergent upon  including more roots whose real parts  lie increasingly far from $\Omega$. 
In Fig.~\ref{Fig2}(c-d), we compare the numerical integration in Eq.~\ref{EQ24_supple} with a restricted sum of the above analytical expression in Eq.~\ref{Eq_GreensFn_Analytical_supple}  for different $d$ values. The contributions of roots in the restricted sum that fall outside the window $p_x\in [\Omega-15\Gamma_{k=0},\Omega+15\Gamma_{k=0}]$ are truncated, which is sufficient to obtain good agreement in the comparison with the numerical results.

Next, we characterize the roots of both the equations $f_{\pm}(p)=0$ and relate them below. We observe in the latter part that the roots contribute to the nonlinear part of the 2-photon transmission current.

\subsubsection{Complex roots of the equations: $f_{\pm}(p)=0.$}\label{App2_giant}

 Here, we prove that the complex roots of the self-energy equation $f_{+}(p)=({p}-\Omega-\Delta_{p})+i\Gamma_{p}/2=0$, which determines the transition frequencies in the dressed-atom, cannot lie in the upper half of the complex plane.
We cast this equation into a pair of coupled nonlinear equations, using $p=p_x+ip_y$, as follows:
\begin{align}
   & p_x-\Omega-2 g_1 g_2 e^{-p_y d}\sin{p_x d}=0,  \label{Non-markov roots giant atom 1}\\
   \text{and}~~& p_y+\big(g_1^2+g_2^2+2 g_1 g_2 e^{-p_y d}\cos{p_x d}\big)=0. \label{Non-markov roots giant atom 2}
\end{align}
Let us suppose there exists a root $(p_x,p_y)$ such that $p_y>0$. We observe $|2 g_1 g_2 e^{-p_y d}\cos{p_x d}| < 2 g_1 g_2 $; since, $e^{-p_y d}$ is a positive, monotonically decreasing function of
$p_y$, and it takes values less than unity for all $p_y>0$. Hence, we get the inequality $(g_1-g_2)^2<g_1^2+g_2^2+2 g_1 g_2 e^{-p_y d}\cos{p_x d}<(g_1+g_2)^2$. This further gives back $p_y<0$ from Eq.~\ref{Non-markov roots giant atom 2}, which is in contradiction to our initial assumption. Therefore, the complex roots of the equation $f_{+}(p)=0$ lie in the lower half of the complex plane. In a similar manner, we can prove that the complex roots of the equation $f_{-}(p)=({p}-\Omega-\Delta_{p})-i\Gamma_{p}/2=0$ cannot lie in the lower half of the complex plane.
It is further clear from the structure of the above equations that if $p_j$ is  a root of the equation $f_{+}(p)=0$, then  $z_j=\Re[p_j]-i\Im[p_j]$ will satisfy  $f_{-}(z_j)=0$.

We emphasize that the roots are countably infinite and all lie in the lower part of the complex plane. Fig.~\ref{Fig2}(a-b) provides a graphical representation of the roots at the intersection of two contour curves taken from the real and imaginary parts of the equation $f_{+}(p)=0$ with $p=p_x+i p_y$. In  Fig.~\ref{Fig2}(a-b), we choose a window $[-3\Gamma_{k=0},3\Gamma_{k=0}]$ around $p_x=\Omega$ in the complex plane $(p_x,p_y)$, and observe that the density of roots  increases  with increasing $d$. The spacing of the roots along $p_x$ will be nearly equal to $2\pi/d$ for large values of $d\gg \lambda$.

\subsubsection{2-photon transmission current}
In the manuscript, we define the 2-photon transmission current (from left to right) as: $J_{L\to R}(2;k)=J^{o}_{L\to R}(2;k)+\delta J_{L\to R}(2;k)$.
The transmission current for two non-interacting photons is defined by
\begin{align}
    J^{o}_{L\to R}(2;k)=\langle\Phi_{2}(R,k;{R},{k})|\hat{J}_{L\to R}(x)|\Phi_{2}(R,k;{R},{k})\rangle~~~\text{for}~~x>d/2.
\end{align}
This contribution constitutes the linear part of the transmission current.
Further,  the contributions from the  nonlinear transmission current are divided into two parts, such as $\delta J_{L\to R}(2;k)=\delta J^{c}_{L\to R}(2;k)+\delta J^{s}_{L\to R}(2;k)$. The cross-correlation component   contains contributions from both the interacting and non-interacting parts of the 2-photon scattering state as follows:
\begin{align}
    \delta J^{c}_{L\to R}(2;k)= 2\Re[\langle\Phi_{2}(R,k;{R},{k})|\hat{J}_{L\to R}(x)|S_{2}(R,k;{R},{k})\rangle]~~~\text{for}~~x>d/2.
\end{align}
The second term in the nonlinear current solely arises from the interacting 2-photon bound state, and it is given by 
\begin{align}
    \delta J^{s}_{L\to R}(2;k)=\langle S_{2}(R,k;{R},{k})|\hat{J}_{L\to R}(x)|S_{2}(R,k;{R},{k})\rangle~~~\text{for}~~x>d/2.
\end{align}
\vspace{0.5 cm}\\
\noindent
\textbf{\textrm{I}.  Non-interacting 2-photon current.}-- Two photons are injected as a product state into the waveguide from the left of the atom. Considering the form of $|\Phi_{2}(R,k;{R},{k})\rangle$ in Eq.~\ref{S21_supple},
we get $\hat{a}_{\beta}(x)|\Phi_{2}(R,k;{R},{k})\rangle=\sqrt{2} \phi_{\beta}^{R}(x,k)|\Phi_{1}(R,k)\rangle$ for $\beta=L, R$. Further simplification leads to the transmission current for two non-interacting photons as follows:
\begin{align}
 J^{o}_{L\to R}(2;k)&= 2\big(|\phi_{R}^{R}(x,k)|^{2} -|\phi_{L}^{R}(x,k)|^{2} \big)\delta(0)=\frac{|t^{R}_{k}|^2}{\pi}\delta(0).
\end{align}
where we use  $\delta$-orthonormalization condition for 1-photon scattering states using 
\begin{align}
    \langle \Phi_{1}(\alpha_1,k_{1})|\Phi_{1}(\alpha_2,k_{2})\rangle= \delta_{\alpha_1,\alpha_2}\delta(k_1-k_2). \label{S34_supple}
\end{align}
\vspace{0.5 cm}\\
\noindent
\textbf{\textrm{II}. Cross-correlation 2-photon current.}--
The expression for the cross-correlation 2-photon current, $\delta J^{c}_{L\to R}(2;k)$ is given by 
\begin{align}
     \delta J^{c}_{L\to R}(2;k) 
     &=-8\Re\Bigg[ \Big\{\phi_{R}^{R*}(x,k)\mathbf{I}_{R}(x,k)-\phi_{L}^{R*}(x,k)\mathbf{I}_{L}(x,k)\Big\} \frac{\phi_{b}^{R}(k)|\phi_{b}^{R}(k)|^2}{\langle b;b|\hat{{G}}^{+}_o(2k)|b;b\rangle}\Bigg]~~~\text{for}~~x>d/2.\label{Eq_crosss}
\end{align}
Here, $\Re [...]$  denotes the real part of a complex number. In the above expression, we use the following definition for the function $\mathbf{I}_{\beta}(x,2k-k')$ where $\beta=L,R$:
\begin{align}
    \mathbf{I}_{\beta}(x,2k-k')= \int  dk'_1 \frac{ \phi_{\beta}^{R}(x,k'_{1}) \phi_{b}^{R*}(k'_{1})+\phi_{\beta}^{L}(x,k'_{1}) \phi_{b}^{L*}(k'_{1})}{2k-(k'_1+k')+i 0^{+}}~~~\text{for}~~x>d/2.
\end{align}
 In order to derive Eq.~\ref{Eq_crosss}, we have used the 2-photon identity in momentum space (see Eq.~\ref{I2_Giant}) and the $\delta$-orthonormalization condition for 1-photon states  given in Eq.~\ref{S34_supple}. Using both, we simplify the following expression, which then leads to Eq.~\ref{Eq_crosss}.
   \begin{align}
   & \langle \Phi_{2} (R,k;{R},{k})|\hat{a}^\dagger_{\beta}(x)\hat{a}_{\beta}(x) \hat{{G}}^{+}_o(2k)|b;b\rangle \nonumber\\
    = &2\sum_{\alpha'_1,\alpha'_2}\int \int dk'_1 dk'_2 \frac{\langle \Phi_{2}(R,k;{R},{k})|\hat{a}^\dagger_{\beta}(x)\hat{a}_{\beta}(x) |\Phi_{1}(\alpha'_1,k'_1)\rangle|\Phi_{1}(\alpha'_2,k'_2)\rangle \big(\phi_{b}^{\alpha'_1}(k'_1)\phi_{b}^{\alpha'_2}(k'_2)\big)^*}{2k-(k'_1+k'_2)+i 0^{+}}=4\phi_{\beta}^{R*}(x,k)\phi_{b}^{R*}(k) \mathbf{I}_{\beta}(x,k).\label{Eq_S76}
\end{align}
Next, we evaluate the integrals in $\mathbf{I}_{L}(x,2k-k')$ and $\mathbf{I}_{R}(x,2k-k')$, for $x>d/2$, without any approximations.
\begin{align}
\textbf{(a) Integration :} ~~~~  \mathbf{I}_{L}(x,2k-k')
    =\frac{1}{2\pi} \int  dk'_1 \frac{ e^{-i k'_{1} x} }{2k-(k'_1+k')+i 0^{+}}\frac{(g_2e^{ik'_{1}d/2}+g_1e^{-ik'_{1}d/2})}{f_{-}(k'_1)}=0,
\end{align}
where we take an integration contour which encloses the lower half of the complex plane.
\begin{align}
 \textbf{(b) Integration :} ~~~~  \mathbf{I}_{R}(x,2k-k') &=\frac{1}{2\pi }\int   \frac{dk'_1~e^{i k'_{1} x} }{2k-(k'_1+ k')+i 0^{+}}\frac{(g_1e^{ik'_{1}d/2}+g_2e^{-ik'_{1}d/2})}{f_{+}(k'_1)}~~~~~~~~~~~~~\nonumber\\
 &=-ie^{i (2k- k') x}  \frac{\big(g_1e^{i(2k-k')d/2}+g_2e^{-i(2k- k')d/2}\big)}{f_{+}(2k-k')}.
\end{align}
In the above line, we take an integration contour that encloses the upper half of the complex plane. Therefore, the 2-photon cross-correlation  current gets simplified to
\begin{align}
    &\delta J^{c}_{L\to R}(2;k) 
     =\frac{2 |\phi_{b}^{R}(k)|^2}{\pi}\Im\Bigg[ \frac{t^{R*}_{k}\Gamma_{k}}{({k}-\Omega-\Delta_{k}+i\Gamma_{k}/2)^2}\frac{1}{\langle b;b|\hat{{G}}^{+}_o(2k)|b;b\rangle}\Bigg],\label{Eq_S93}
\end{align}
which is invariant under the transformation $g_1\leftrightarrow g_2$. Here, $\Im[...]$  denotes the imaginary part of a complex number. Within the Markov approximation, the expression with the braces in Eq.~\ref{Eq_S93} becomes
\begin{align}
    \Bigg[ \frac{t^{R*}_{k}\Gamma_{k}}{({k}-\Omega-\Delta_{k}+i\Gamma_{k}/2)^2}\frac{1}{\langle b;b|\hat{{G}}^{+}_o(2k)|b;b\rangle}\Bigg]\approx \Bigg[ \frac{(k-\Omega-\Delta_{\Omega})\Gamma_{\Omega} ({k}-\Omega-\Delta_{\Omega}+i\Gamma_{\Omega}/2)}{({k}-\Omega-\Delta_{\Omega}+i\Gamma_{\Omega}/2)^2 ({k}-\Omega-\Delta_{\Omega}-i\Gamma_{\Omega}/2)} \Bigg]=0.
\end{align}
and we get $\delta J^{c}_{L\to R}(2;k)\approx 0$ in the Markov regime.
\vspace{1.0 cm}\\
\noindent
\textbf{\textrm{III}. Solely bound-state contribution to 2-photon current.}--
The nonlinear transmission current, which solely arises due to the interacting 2-photon bound state, is given by
\begin{align}
    \delta J^{s}_{L\to R}(2;k)= 4\langle b;b|\hat{{G}}^{-}_o(2k) \hat{J}_{L \to R}(x) \hat{{G}}^{+}_o(2k)|b;b\rangle \frac{|\phi_{b}^{R}(k)|^4}{|\langle b;b|\hat{{G}}^{+}_o(2k)|b;b\rangle|^2}.
\end{align}
where $\hat{{G}}^{-}_o(2k)=[\hat{{G}}^{+}_o(2k)]^\dagger$.
To simplify the above expression, we need $\langle b;b|\hat{{G}}^{-}_o(2k) \hat{a}^\dagger_{\beta}(x)\hat{a}_{\beta}(x)\hat{{G}}^{+}_o(2k)|b;b\rangle$ where $\beta=R,L$. We use 2-photon identity resolution in momentum space (see Eq.~\ref{I2_Giant}) and get
\begin{align}
    &\hat{a}_{\beta}(x) \hat{{G}}^{+}_o(2k)|b;b\rangle =2\sqrt{2}\sum_{\alpha'_1,\alpha'_2}\int \int dk'_1 dk'_2~\frac{|\Phi_{1}(\alpha'_{2},k'_{2})\rangle \phi_{\beta}^{\alpha'_{1}}(x,k'_{1}) }{2k-(k'_1+k'_2)+i 0^{+}} \big(\phi_{b}^{\alpha'_{1}}(k'_{1})\phi_{b}^{\alpha'_{2}}(k'_{2})\big)^{*}.
    \end{align}
Using the above expression, we  can derive
\begin{align}
    &
    \langle b;b|\hat{{G}}^{-}_o(2k) \hat{a}^\dagger_{\beta}(x)\hat{a}_{\beta}(x) \hat{{G}}^{+}_o(2k)|b;b\rangle
   =8\sum_{\alpha'}\int  dk' |\phi_{b}^{\alpha'}(k')\mathbf{I}_{\beta}(x,2k-k')|^{2}.
\end{align}
Considering $x>d/2$, we finally have
\begin{align}
   \delta J^{c}_{L\to R}(2;k) 
   =32\pi \frac{|\phi_{b}^{R}(k)|^4}{|\langle b;b|\hat{{G}}^{+}_o(2k)|b;b\rangle|^2}\int  dk' |\phi_{b}^{R}(k') \phi_{b}^{R}(2k-k')|^{2}=\frac{|\phi^{R}_b(k)|^4}{|\langle b;b|\hat{G}_{o}^{+}(2k)|b;b\rangle|^2} \mathcal{T}(2k). \label{Eq_S98}
\end{align}
Both the pre-factor and the integrand are invariant under the transformation $g_1 \leftrightarrow g_2$. Hence, the current in Eq.~\ref{Eq_S98} is likewise invariant under $g_1 \leftrightarrow g_2$.
In the following steps, we evaluate $\mathcal{T}(2k)$ using the residue theorem.
\begin{align}
    \mathcal{T}(2k) 
    &=\frac{2}{\pi}\int  dk' \frac{\Gamma_{k'}}{f_{+}(k')f_{-}(k')}\frac{\Gamma_{2k-k'}}{f_{-}(2k-k')f_{+}(2k-k')}.
\end{align}
We  enclose the integration  contour in the lower half of the complex plane. The equations governing poles in the upper plane are: $f_{+}(k')=0$, and $f_{-}(2k-k')=0$ (see the discussions in Subsection~\ref{App2_giant}). Here, $f'_{-}(2k-k')=-1+g_1g_2 d e^{-i(2k-k') d}$.
Therefore, the roots $\{p_j\}$ of $f_{+}(k')=0$ and the roots $\{q_j\}$ of $f_{-}(2k-k')=0$, contribute the residues as simple poles. After simplification, we obtain
\begin{align}
    \mathcal{T}(2k)=-&4i\sum_j\Bigg[\frac{\Gamma_{p_j}\Gamma_{2k-p_j}}{f'_{+}(p_j)f_{-}(p_j) f_{+}(2k-p_j)f_{-}(2k-p_j)}+\frac{\Gamma_{q_j}\Gamma_{2k-q_j}}{f_{+}(q_j)f_{-}(q_j)f'_{-}(2k-q_j)f_{+}(2k-q_j)}\Bigg].
\end{align}
Following the discussions in Subsection~\ref{App2_giant}, we can show that the roots $\{q_j\}$ of the equation $f_{-}(2k-k')=0$ are related to $\{p_j\}$ as follows: $q_j=2k-\Re[p_j]+\Im[p_j]$ where all $j\in \mathbb{Z}^+$. Under the Markov approximation, we get 
\begin{gather}
  \mathbf{I}_{R}(x,2k-k') \approx -ie^{i (2k- k') x}  \frac{(g_1e^{i\Omega d/2}+g_2e^{-i\Omega d/2})}{(2k-k')-\Omega-\Delta_{\Omega}+i\Gamma_{\Omega}/2},\\
 \mathcal{T}(2k)\approx\frac{8\Gamma_\Omega}{(2k-2\Omega-2\Delta_{\Omega})^2+\Gamma_{\Omega}^2} ~~~\text{and}~~~  \delta J^{c}_{L\to R}(2;k) \approx   \frac{\Gamma_{\Omega}^3 }{2(2\pi)^2\big[({k}-\Omega-\Delta_{\Omega})^2+\Gamma^2_{\Omega}/4\big]^2}. \label{MArkov_Gianttwophoton_intcurrent}
\end{gather}
Therefore, we demonstrate that  all nonlinear current components are invariant under $g_1 \leftrightarrow g_2$. Hence, the giant atom-waveguide setup gives reciprocal 2-photon transport, in spite of its inherent structural asymmetry when $g_1 \neq g_2$.

\section{Generalized direct-coupled atom-waveguide setup}
In the manuscript, we discuss a generalized  direct-coupled WQED setup \cite{Segal_PRL_sufficient, DRoy_OpticalDiode_PRB2010} where transport properties for correlated photons can be tuned from the nonreciprocal to reciprocal regimes in parity-broken configurations. In this setup, an atom is directly coupled to two semi-infinite waveguides at both ends, plus an additional tunneling path exists for the photons coupling the two waveguides. The Hamiltonian of such a system can be written as 
\begin{align}   
    \hat{H}^d =\hat{H}_b&+\hat{H}'_{\text{w}}+\Big[\big(g_L \hat{a}^{\dagger}_{1}(0)+g_R \hat{a}^{\dagger}_{2}(0)\big)\hat{b}+\text{H.c.}\Big]+2J \hat{c}^\dagger_{1}(0)\hat{c}_{2}(0)+2J^*\hat{a}^\dagger_{2}(0)\hat{a}_{1}(0),\label{Hamiltonian_dcouple_Supple}
\end{align}
where the Hamiltonian for the waveguides is given by (with $v_g=1$)
\begin{align}
    \hat{H}'_{\text{w}}=-i&\int_{-\infty}^{\infty} dx \Big[\hat{a}^{\dagger}_{1}(x)\partial_{x} \hat{a}_{1}(x)+\hat{a}^{\dagger}_{2}(x)\partial_{x} \hat{a}_{2}(x)\Big].
\end{align}
The tunneling amplitude $2J$ in the Hamiltonian controls direct coupling of photons between channel-$1$ and channel-$2$,  without involving the atom. This  mimics a regulated interference path in the setup without a time lag. The atom is coupled to the channel-1 and channel-2 with amplitudes $g_L$ and $g_R$.
We consider $g_L, g_R,J$ to be complex in general. 
In the following analysis, we obtain the transmission current operator in this setup and study the  transport with  1-photon and 2-photon input states in the waveguides when $|g_L|\neq |g_R|$.

\subsection{Transmission  current operator}
Similar to the giant atom setup, we define local density operator in the channel-$l$  by  $\hat {n}_{l}(x)= \hat{a}^\dagger_{l}(x)\hat{a}_{l}(x)+\hat{a}^\dagger_{l}(-x)\hat{a}_{l}(-x)$, where $l=1,2$. Using the Heisenberg equations of motion for the field operators, we obtain
the rate of change $\hat n_{l} (x)$ with time $t$ as follows:
\begin{align}
    &\frac{\partial }{\partial t} \hat n_{l}(x)=-\frac{\partial}{\partial x}\Big[ \hat{a}^\dagger_{l}(x)\hat{a}_{l}(x)-\hat{a}^\dagger_{l}(-x)\hat{a}_{l}(-x)\Big]~~~\text{for}~~x\neq 0.
\end{align}
Therefore,  the transmission current operator in the channel-$l$ is $  \hat{J_l}(x)=[\hat{a}^\dagger_{l}(x)\hat{a}_{l}(x)-\hat{a}^\dagger_{l}(-x)\hat{a}_{l}(-x)]$ for $x>0$.
When photons are injected in the channel-1 from the left side of the atom, we use $\hat{J_2}(x>0)$ to obtain the average transmission photon current. Similarly, $\hat{J_1}(x>0)$ gives the transmission current when photons are injected in the channel-2 from the right side of the atom.

\subsection{1-photon paradigm}
Scattering states of $\hat{H}^d$ in the single-excitation sector are denoted by
\begin{align}
|\tilde{\Phi}_{1}(\alpha,k)\rangle=&\int_{-\infty}^{\infty} dx \big[\tilde{\phi}_{1}^\alpha(k,x) a^\dagger_{1}(x)+\tilde{\phi}_{2}^\alpha(k,x)a^\dagger_{2}(x)\big]|\varphi;0\rangle+\tilde{\phi}_{b}^{\alpha}(k) b^\dagger|\varphi;0\rangle,\label{Single_photon_waveFunction}
\end{align}
 Similar to the giant atom case, the superscript index $\alpha(=L, R)$ denotes the left- or right-moving photon in the initial state with a momentum $k$. We use the time-independent Schr\"{o}dinger  equation and get 
\begin{align}
  & \big(\Omega-{k}\big)\tilde{\phi}_{b}^{\alpha}(k)+\Big[g^{*}_L\tilde{\phi}_{1}^{\alpha}(k,0)+g^{*}_R\tilde{\phi}_{2}^{\alpha}(k,0)\Big] =0,\label{Eq_SinglePhoton_Mapped_emitter}\\
  & \big(i\partial_x+{k}\big)\tilde{\phi}_{1}^{\alpha}(k,x)=g_L\delta(x)\tilde{\phi}_{b}^{\alpha}(k)+J\delta(x)\tilde{\phi}_{2}^\alpha(k,0),\label{Eq_SinglePhoton_Mapped_left}\\
  & \big(i\partial_x+{k}\big)\tilde{\phi}_{2}^{\alpha}(k,x)=g_R\delta(x)\tilde{\phi}_{b}^{\alpha}(k)+J^{*}\delta(x)\tilde{\phi}_{1}^\alpha(k,0).
\end{align}

{{\textbf{ \textrm{I}. Incident photon in channel-1}}}.--
We first consider a single right-moving photon incident from the left of the atom, and propagating with a momentum $k$. The scattering ansatz for the wavefunction amplitudes is
\begin{align} 
\tilde{\phi}_{1}^{R}(k,x)=\frac{e^{ikx}}{\sqrt{2\pi}}\big[\Theta(-x)+\tilde{r}^{R}_k\Theta(x)\big],~~~
\tilde{\phi}_{2}^{R}(k,x)=\frac{e^{ikx}}{\sqrt{2\pi}} \tilde{t}^{R}_k\Theta(x).
\end{align}
In the above equation, the reflection and transmission amplitudes  for the right-moving photons  are denoted by $\tilde{t}_k^{R}$ and $\tilde{r}_k^{R}$, respectively. 
After simplification, we get
\begin{gather}
  \tilde{\phi}_{b}^{R}(k)=\sqrt{\frac{1}{2\pi}}\frac{  ( g_L^*-iJ^*g_R^*)/(|J|^2+1)}{({k}-\Omega-\Delta)+i\Gamma/2},\\
  \tilde{t}^{R}_{k}=\frac{  -i[2J^* ({k}-\Omega)+ g_Rg_L^* ]/(|J|^2+1)}{({k}-\Omega-\Delta)+i\Gamma/2},~~~
    \tilde{r}^{R}_{k}=-\frac{\Big[\frac{(|J|^2-1)}{(|J|^2+1)}({k}-\Omega)+i\frac{(|g_L|^2-|g_R|^2)}{2(|J|^2+1)}\Big]-\Delta}{({k}-\Omega-\Delta)+i\Gamma/2},
\end{gather}
The Lamb shift $(\Delta)$ and the relaxation rate $(\Gamma)$ are given by 
\begin{align}
    \Delta=-\frac{(g_L g_R^* J^*+g_L^* g_R J)}{2(|J|^2+1)},~~~
    \Gamma=\frac{(|g_L|^2+|g_R|^2)}{(|J|^2+1)}.
\end{align}
We observe that the generalized direct-coupled atom in a linear waveguide acquires a finite Lamb shift whenever $J\neq 0$.

{{\textbf{ \textrm{II}. Incident photon in channel-2}}}.--
The scattering ansatz for the left-moving photon is:
\begin{align} 
\tilde{\phi}_{1}^{L}(k,x)&=\frac{ e^{ikx}}{\sqrt{2\pi}}\tilde{t}^{L}_{k} \Theta(x),~~~~
\tilde{\phi}_{2}^{L}(k,x)=\frac{e^{ikx}}{\sqrt{2\pi}}\big[ \Theta(-x)+\tilde{r}^{L}_k \Theta(x)\big],
\end{align}
where the reflection and transmission amplitudes are $\tilde{t}_k^{L}$ and $\tilde{r}_k^{L}$, respectively. 
Simplification leads to
\begin{gather}
 \tilde{\phi}_{b}^{L}(k)=\sqrt{\frac{1}{2\pi}}\frac{  ( g_R^*-iJg_L^*)/(|J|^2+1)}{({k}-\Omega-\Delta)+i\Gamma/2},\\
 \tilde{t}^{L}_{k}=\frac{  -i[2J ({k}-\Omega)+ g_Lg_R^* ]/(|J|^2+1)}{({k}-\Omega-\Delta)+i\Gamma/2},~~~
   \tilde{r}^{L}_{k}=-\frac{\Big[\frac{(|J|^2-1)}{(|J|^2+1)}({k}-\Omega)-i\frac{(|g_L|^2-|g_R|^2)}{2(|J|^2+1)}\Big]-\Delta}{({k}-\Omega-\Delta)+i\Gamma/2}.
\end{gather}
Therefore, we get $|\tilde{t}^{R}_{k}|^2=|\tilde{t}^{L}_{k}|^2$, i.e., the transmission probability of a single photon from the left to right is the same as that of a photon from the right to left.  On the other hand, the excitation probability of the atom in general depends on the propagation direction of the incident photon. 
It is interesting to consider the special limit $g_R=g_L=0$. This yields $\tilde{\phi}_{b}^{\alpha}(k)=0$, $\tilde{t}^{R}_k=-2iJ^{*}/(|J|^2+1)$, and $\tilde{t}^{L}_k=-2iJ/(|J|^2+1)$. Hence, $J=\pm i,\pm 1$ ensures perfect transmission of single photons between the waveguides when the atom is decoupled.

\subsubsection*{ Mapping relations between the giant atom setup and the generalized direct-coupled setup}
We consider $J=-i$ in the generalized direct-coupled model. The excitation amplitude of the atom, the transmission and the reflection amplitudes of the scattered 1-photon state for a right-incident photon are given by
\begin{gather}
  \tilde{\phi}_{b}^{R}(k)=\sqrt{\frac{1}{2\pi}}\frac{  ( g_L^*+g_R^*)/2}{({k}-\Omega-\Delta)+i\Gamma/2},~~~
  \tilde{t}^{R}_{k}=\frac{ ({k}-\Omega)-i g_Rg_L^* /2}{({k}-\Omega-\Delta)+i\Gamma/2},~~~
    \tilde{r}^{R}_{k}=-\frac{i(|g_L|^2-|g_R|^2)/4-\Delta}{({k}-\Omega-\Delta)+i\Gamma/2},
\end{gather}
where the Lamb shift and the relaxation rate become $\Delta=-i(g_L g_R^* -g_L^* g_R )/4,$ and $ \Gamma=(|g_L|^2+|g_R|^2)/2$.
Since, the coupling amplitudes are complex parameters, we choose $g_L=2(x+i y)$ and $g_R=2(v+i w).$
To initiate a mapping, we put the following constraints on the coupling amplitudes:
\begin{gather}
  i g_L^* g_R /2=2g_1g_2 \sin{\theta},~~
  (g_L^*+g_R^* )/2=(g_1e^{-i\theta/2}+g_2e^{i\theta/2}),\\
   i(|g_L|^2+|g_R|^2)/4+i(g_L g_R^* -g_L^* g_R)/4=i(g_1^2+g_2^2+2g_1g_2\cos{\theta})-2g_1g_2\sin{\theta}.
\end{gather}
The above constraints give
\begin{align}
    \frac{x}{y}=-\frac{w}{v}=p,~~~(p^2+1)yv= g_1g_2\sin{\theta},~~~
    (p y+v)=(g_1+g_2)\cos{\theta/2},~~~
    (y -p v)=(g_1-g_2)\sin{\theta/2}.
\end{align}
The constraints further give: $\Delta=i g^{*}_L g_R/2,$   and $\Gamma/2=x^2+y^2+v^2+w^2$.
Using the above equations, we find out the following solutions for $g_L=2(x+iy)$ and $g_R=2(v+iw)$:
\begin{gather}
   g_L=\frac{e^{i {\theta}/2}}{g_1+g_2e^{i {\theta}}}\Big[g_1^2+g_2^2+2g_1g_2e^{i{\theta}}+\sqrt{(g_1^2+g_2^2)^2+4g_1g_2\{(g_1^2+g_2^2)\cos{{\theta}}+g_1g_2\cos{2{\theta}}\}}\Big], \label{Eq_gLTrans_Supple}\\
     g_R=\frac{e^{i {\theta}/2}}{g_1+g_2e^{i {\theta}}}\Big[g_1^2+g_2^2+2g_1g_2e^{-i\theta}-\sqrt{(g_1^2+g_2^2)^2+4g_1g_2\{(g_1^2+g_2^2)\cos{{\theta}}+g_1g_2\cos{2\theta}\}}\Big].\label{Eq_gRTrans_Supple}
\end{gather}
 With these choices of the parameters, we get: $\tilde{t}_k^{R}=-\tilde{t}_k^{L}$, $|\tilde{r}_k^{R}|=|\tilde{r}_k^{L}|$ and $|\tilde{\phi}_{b}^{R}(k)|=|\tilde{\phi}_{b}^{L}(k)|$. One can further show $|g_R\pm g_L|^2=2\Gamma$.

Next, if we identify the free parameter $\theta$ in the above transformations  as $\theta=kd$, it guarantees the following mapping relations between the 1-photon amplitudes in the giant atom setup to those in the generalized direct-coupled setup: $ \tilde{t}^{R}_k={t}^{R}_k$, $ \tilde{\phi}^{R}_b(k)= {\phi}^{R}_b(k)$, and $|  \tilde{r}^{R}_k|=|{r}^{R}_k|$. In the same spirit, $\theta=\Omega$ maps the 1-photon amplitudes in the generalized direct-coupled setup to those in the giant atom setup within the Markov approximation. 

\subsection{2-photon paradigm}
Similar to the giant atom case, we use the Lippmann-Schwinger formalism to separate the  2-photon scattering state  $|\tilde{\Psi}_{2}(\alpha,k;\alpha,k)\rangle$ into non-interacting and interacting parts as follows:
\begin{align}
|\tilde{\Psi}_2(\alpha,k;\alpha,k)\rangle=&|\tilde{\Phi}_2(\alpha,k;\alpha,k)\rangle-\hat{{G}}^{+}_o(2k)|b;b\rangle \frac{\langle b;b|\tilde{\Phi}_{2}(\alpha,k;\alpha,k)\rangle}{\langle b;b|\hat{\tilde{G}}^{+}_o(2k)|b;b\rangle}=|\tilde{\Phi}_{2}(\alpha,k;\alpha,k)\rangle+| {\tilde S}_{2}(\alpha,k;\alpha,k)\rangle,\label{Lippmann_direct}
\end{align}
where the photons are injected with the same momentum $k$ and they are either propagating towards the right $(\alpha=R)$, or towards the left $(\alpha=L)$.
As before,   $\hat{\tilde{G}}^{+}_o(2k)=(2k-\hat{H}^{d}_o+i0^{+})^{-1}$ is  the retarded Green's function operator of non-interacting Hamiltonian $\hat{H}^{d}_o=\hat{H}^{d}-U\hat{b}^\dagger\hat{b}^\dagger \hat{b} \hat{b}/2$.   The 2-photon non-interacting state is given by $|\tilde{\Phi}_{2}(\alpha,k;\alpha,k)\rangle=\frac{1}{\sqrt{2}}|\tilde{\Phi}_{1}(\alpha,k)\rangle |\tilde{\Phi}_{1}(\alpha,k)\rangle$,
where the incident photons are injected as a product state into the waveguide.

 \subsubsection{2-photon Green's functions}
 In this subsection, we obtain analytical expressions for the 2-photon Green's functions. They will be later utilized in the 2-photon current and the scattering matrix calculations.
To proceed, we establish a 2-photon identity $\tilde{\mathds{I}}_2$ in the momentum space as follows:
\begin{align}
    \tilde{\mathds{I}}_2=\sum_{\alpha'_1,\alpha'_2} \int \int dk'_1 dk'_2~|\tilde{\Phi}_{1}(\alpha'_1,k'_1)\rangle  |\tilde{\Phi}_{1}(\alpha'_2,k'_2)\rangle \langle\tilde{\Phi}_{1}(\alpha'_1,k'_1)|  \langle\tilde{\Phi}_{1}(\alpha'_2,k'_2)|. \label{I2_direct}
\end{align}
Evaluating various 2-photon integrations using complex analysis, we obtain the following Green's functions: 
\begin{align}
 &\langle b;b|\hat{\tilde{G}}^{+}_o(E_{\mathbf k})|b;b\rangle
    =\frac{2}{E_{\mathbf k}-2(\Omega+\Delta)+i\Gamma},\\
&\langle x_1,1;x_2,1| \hat{\tilde{G}}^{+}_o(E_{\mathbf k})|b;b\rangle=-2\Theta(x_1)\Theta(x_2)\bigg[\frac{(g_L-i J g_R )}{|J|^2+1}\bigg]^2\frac{e^{i E_{\mathbf k} x_c}e^{i (E_{\mathbf k}/2-\Omega-\Delta+i\Gamma/2)|x|}}{E_{\mathbf k}-2(\Omega+\Delta)+i \Gamma} , \\
&\langle x_1,2;x_2,2| \hat{\tilde{G}}^{+}_o(E_{\mathbf k})|b;b\rangle=-2\Theta(x_1)\Theta(x_2)\bigg[\frac{(g_R-i J^{*} g_L )}{|J|^2+1}\bigg]^2\frac{e^{i E_{\mathbf k} x_c}e^{i (E_{\mathbf k}/2-\Omega-\Delta+i\Gamma/2)|x|}}{E_{\mathbf k}-2(\Omega+\Delta)+i \Gamma} ,\\
&\langle x_1,1;x_2,2| \hat{\tilde{G}}^{+}_o(E_{\mathbf k})|b;b\rangle=-2\sqrt{2} \Theta(x_1)\Theta(x_2)\bigg[\frac{(g_R-i J^{*} g_L )}{(|J|^2+1)} \frac{(g_L-i J g_R )}{(|J|^2+1)}\bigg]\frac{ e^{i E_{\mathbf k} x_c}e^{i (E_{\mathbf k}/2-\Omega-\Delta+i\Gamma/2)|x|}}{E_{\mathbf k}-2(\Omega+\Delta)+i \Gamma}.
\end{align}
In the above calculations, 
we use $|x_1,l;x_2,l\rangle= \hat{a}^\dagger_{l}(x_1) \hat{a}^\dagger_{l}(x_2)|\varphi;0\rangle/\sqrt{2}$ for channels $l=1,2$, along with the following relation $|x_1,1;x_2,2\rangle= \hat{a}^\dagger_{1}(x_1) \hat{a}^\dagger_{2}(x_2)|\varphi;0\rangle.$ Further, we denote the total energy of two photons as  $E_{\mathbf{k}}=k_1+k_2$, the relative 2-photon coordinate as $x=x_1-x_2$, and the center of mass coordinate by $x_c=(x_1+x_2)/2$.

\subsubsection{2-photon transmission current}
As mentioned in the manuscript, our goal is to obtain 2-photon transmission current in this generalized direct-coupled setup to demonstrate a transition from nonreciprocal to reciprocal regimes in the parity-broken configuration $|g_L|\neq |g_R|$.
Similarly to the giant atom case, the average 2-photon transmission current from the left to right can be separated into different parts as follows:  $\tilde{J}_{L\to R}(2;k)=\tilde{J}^{o}_{L\to R}(2;k)+\delta \tilde{J}_{L\to R}(2;k)$, where $\delta \tilde{J}_{L\to R}(2;k)=\delta \tilde{J}^{c}_{L\to R}(2;k)+\delta \tilde{J}^{s}_{L\to R}(2;k)$. The components are given by
\begin{align}
    &\tilde{J}^{o}_{L\to R}(2;k)=\langle\tilde{\Phi}_{l}(R,k;R,k)|\hat{J}_{l}(x)|\tilde{\Phi}_{2}(R,k;R,k)\rangle,\\
    &\delta \tilde{J}^{c}_{L\to R}(2;k)= 2\Re[\langle \tilde{\Phi}_{2}(R,k;R,k)|\hat{J}_{l}(x)|\tilde{S}_{2}(R,k;R,k)\rangle],\\
    &\delta \tilde{J}^{s}_{L\to R}(2;k)=\langle \tilde{S}_{2}(R,k;R,k)|\hat{J}_{l}(x)|\tilde{S}_{2}(R,k;R,k)\rangle,
\end{align}
where $l=2$ and $x>0$ in the above expressions. Now, the transmission current components, from right to left ($R \to L$) across the atom, are obtained by replacing $R$ with $L$ and setting $l=1$ in the above expressions. 
\vspace{0.5 cm}\\
\noindent
\textbf{\textrm{I}.  Non-interacting 2-photon current}.--
We first  consider  the linear term in the 2-photon current due to non-interacting photons. 
We utilize $\hat{a}_{l}(x)|\tilde{\Phi}_{2}(R,k;R,k)\rangle=\sqrt{2} \tilde{\phi}_{l}^{R}(x,k)|\tilde{\Phi}_{1}(R,k)\rangle$ for channels $l=1, 2$. After simplification, we get
\begin{align}
 &\tilde{J}^{o}_{L\to R}(2;k)= 2\big(|\tilde{\phi}_{2}^{R}(x,k)|^{2} -|\tilde{\phi}_{2}^{R}(-x,k)|^{2} \big)\delta(0)=\frac{|\tilde{t}^{R}_{k}|^2}{\pi}\delta(0),\\
 &\tilde{J}^{o}_{R\to L}(2;k)= 2\big(|\tilde{\phi}_{2}^{L}(x,k)|^{2} -|\tilde{\phi}_{2}^{L}(-x,k)|^{2} \big)\delta(0)=\frac{|\tilde{t}^{L}_{k}|^2}{\pi}\delta(0).
\end{align}
Here, we use  $\delta$-orthonormalization condition on the 1-photon scattering states $|\tilde{\Phi}_{1}(\alpha,k)\rangle$ for $\alpha=R,L$.
\vspace{0.5 cm}\\
\noindent
\textbf{\textrm{II}. Cross-correlation 2-photon current}.--
We obtain the cross-correlation part of the 2-photon transmission current, $\delta J^{c}_{L\to R}(2;k)$. The cross-correlation part arises from the contribution of both non-interacting 2-photon state and interacting 2-photon bound state. We derive
\begin{align}
     \delta \tilde{J}^{c}_{L\to R}(2;k) 
     &=-8\Re\Bigg[ \Big\{\tilde{\phi}_{2}^{R*}(x,k)\tilde{\mathbf{I}}_{2}(x,k)-\tilde{\phi}_{2}^{R*}(-x,k)\tilde{\mathbf{I}}_{2}(-x,k)\Big\} \frac{\tilde{\phi}_{b}^{R}(k)|\tilde{\phi}_{b}^{R}(k)|^2}{\langle b;b|\hat{\tilde{G}}^{+}_o(2k)|b;b\rangle}\Bigg]~~\text{for}~x>0,\label{Eq_crossDirect_LR}\\
     \delta \tilde{J}^{c}_{R\to L}(2;k) 
     &=-8\Re\Bigg[ \Big\{\tilde{\phi}_{1}^{L*}(x,k)\tilde{\mathbf{I}}_{1}(x,k)-\tilde{\phi}_{1}^{L*}(-x,k)\tilde{\mathbf{I}}_{1}(-x,k)\Big\} \frac{\tilde{\phi}_{b}^{L}(k)|\tilde{\phi}_{b}^{L}(k)|^2}{\langle b;b|\hat{\tilde{G}}^{+}_o(2k)|b;b\rangle}\Bigg]~~\text{for}~x>0.\label{Eq_crossDirect_RL}
\end{align}
In the above derivation, we use $\langle \tilde{\Phi}_{2} (\alpha,k;\alpha,k)|\hat{a}^\dagger_{l}(x)\hat{a}_{l}(x) \hat{\tilde{G}}^{+}_o(2k)|b;b\rangle=4\phi_{l}^{\alpha*}(x,k)\phi_{b}^{\alpha*}(k) \tilde{\mathbf{I}}_{l}(x,k)$, where
\begin{align}
     \tilde{\mathbf{I}}_{l}(x,2k-k')= \int  dk'_1 \frac{ \tilde{\phi}_{l}^{R}(x,k'_{1}) \tilde{\phi}_{b}^{R*}(k'_{1})+\tilde{\phi}_{l}^{L}(x,k'_{1}) \tilde{\phi}_{b}^{L*}(k'_{1})}{2k-(k'_1+k')+i 0^{+}}~~~\text{where}~~x>0,
\end{align}
and for channels $l=1,2$. 
We evaluate the integrations in $\tilde{\mathbf{I}}_{2}(\pm x,2k-k')$ and $\tilde{\mathbf{I}}_{1}(\pm x,2k-k')$ for $x>0$ as follows:
\begin{align}
&\textbf{(a) Integration :} ~~~~ \tilde{\mathbf{I}}_{2}(-x,2k-k')=\frac{1}{2\pi} \int  \frac{ dk'_1~ e^{-ik'_1x} }{2k-(k'_1+k')+i 0^{+}}\frac{( g_R+iJ^*g_L)/(|J|^2+1)}{(k'_1-\Omega-\Delta)-i\Gamma/2}=0, \\
&\textbf{(b) Integration :} ~~   
  \tilde{\mathbf{I}}_{2}(x,2k-k') 
     =\frac{1}{2\pi}\int \frac{ dk'_1~e^{ik'_1x} }{2k-(k'_1+k')+i 0^{+}}\frac{(g_R-i g_L J^*)/(|J|^2+1)}{({k'_1}-\Omega-\Delta)+i\Gamma/2}=-ie^{i(E-k_2)x} \frac{(g_R-i g_L J^*)/(|J|^2+1)}{(2k-{k'}-\Omega-\Delta)+i\Gamma/2},\\
 & \textbf{(c) Integration :} ~~~~~\tilde{ \mathbf{I}}_{1}(-x,2k-k')=\frac{1}{2\pi} \int  \frac{ dk'_1~ e^{-ik'_1x} }{2k-(k'_1+k')+i 0^{+}}\frac{( g_L+iJg_R)/(|J|^2+1)}{(k'_1-\Omega-\Delta)-i\Gamma/2}=0,\\
&\textbf{(d) Integration :} ~~   
  \tilde{\mathbf{I}}_{1}(x,2k-k') 
     =\frac{1}{2\pi}\int \frac{ dk'_1~e^{ik'_1x} }{2k-(k'_1+k')+i 0^{+}}\frac{(g_L-i g_R J)/(|J|^2+1)}{({k'_1}-\Omega-\Delta)+i\Gamma/2}=-ie^{i(E-k_2)x} \frac{(g_L-i g_R J)/(|J|^2+1)}{(2k-{k'}-\Omega-\Delta)+i\Gamma/2}.
\end{align} 
Therefore, we simplify the expression for the cross-correlation 2-photon currents and  obtain
\begin{align}
    &\delta \tilde{J}^{c}_{L\to R}(2;k) 
     =\frac{4}{\pi}\Im\Bigg[ \frac{\tilde{t}^{R*}_{k}  (g_R-i g_L J^*)( g_L^*-iJ^*g_R^*)}{({k}-\Omega-\Delta+i\Gamma/2) (|J|^2+1)^2}|\tilde{\phi}_{b}^{R}(k)|^2\Bigg],\\
      &\delta \tilde{J}^{c}_{R\to L}(2;k) 
     =\frac{4}{\pi}\Im\Bigg[ \frac{\tilde{t}^{L*}_{k}  (g_L-i g_R J)( g_R^*-iJg_L^*)}{({k}-\Omega-\Delta+i\Gamma/2) (|J|^2+1)^2}|\tilde{\phi}_{b}^{L}(k)|^2\Bigg].
\end{align}
We observe $\delta \tilde{J}^{c}_{L\to R}(2;k) \neq \delta \tilde{J}^{c}_{R\to L}(2;k)$ for the generic cases when $|\tilde{\phi}_{b}^{R}(k)|\neq |\tilde{\phi}_{b}^{L}(k)|$. 

However, (a) taking all parameters $g_L,g_R,J$ to be reals with $J=\pm 1$, we see  $\delta \tilde{J}^{c}_{L\to R}(2;k) = \delta \tilde{J}^{c}_{R\to L}(2;k)$ while keeping $g_L \neq g_R$.
Next, (b) choosing the coupling parameters according to Eq.~(\ref{Eq_gLTrans_Supple}-\ref{Eq_gRTrans_Supple}) with $J=-i$, we get 
\begin{align}
    &\delta \tilde{J}^{c}_{L\to R}(2;k) 
     =\frac{2}{\pi}\Im\Bigg[ \frac{\tilde{t}^{R*}_{k}  \Gamma}{({k}-\Omega-\Delta+i\Gamma/2)}|\tilde{\phi}_{b}^{R}(k)|^2\Bigg], ~~\text{and}~~
      \delta \tilde{J}^{c}_{R\to L}(2;k) 
     =\frac{2}{\pi}\Im\Bigg[ \frac{-\tilde{t}^{L*}_{k} \Gamma}{({k}-\Omega-\Delta+i\Gamma/2)}|\tilde{\phi}_{b}^{L}(k)|^2\Bigg].\label{expression}
\end{align}
Since, the parameter choices further give $\tilde{t}^{R}_{k}=-\tilde{t}^{L}_{k}$, and $|\tilde{\phi}_{b}^{R}(k)|^2=|\tilde{\phi}_{b}^{L}(k)|^2$, we get $\delta \tilde{J}^{c}_{L\to R}(2;k)=\delta \tilde{J}^{c}_{L\to R}(2;k)=0$. In both cases, the term inside the brackets is purely real.
\vspace{0.5 cm}\\
\noindent
\textbf{\textrm{III}. Solely bound-state contribution to 2-photon current}.--
Next, we obtain the part of the nonlinear transmission current, $\delta J^{s}_{L\to R}(2;k)$, solely arising from the interacting 2-photon bound state. For right- and left-incident photons, we get the following expressions
\begin{align}
   & \delta {\tilde J}^{s}_{L\to R}(2;k)=4 \langle b;b|\hat{\tilde{G}}^{-}_o(2k) \hat{J}_{2}(x) \hat{{G}}^{+}_o(2k)|b;b\rangle \frac{|\tilde{\phi}_{b}^{R}(k)|^4}{|\langle b;b|\hat{\tilde{G}}^{+}_o(2k)|b;b\rangle|^2}~~~\text{for}~x>0,\\
     &\delta {\tilde J}^{s}_{L\to R}(2;k)= 4\langle b;b|\hat{\tilde{G}}^{-}_o(2k) \hat{J}_{1}(x) \hat{{G}}^{+}_o(2k)|b;b\rangle \frac{|\tilde{\phi}_{b}^{L}(k)|^4}{|\langle b;b|\hat{\tilde{G}}^{+}_o(2k)|b;b\rangle|^2}~~~\text{for}~x>0,
\end{align}
where $\hat{{G}}^{-}_o(2k)=[\hat{{G}}^{+}_o(2k)]^\dagger$. Using  the 2-photon identity resolution in momentum space, we get
\begin{align}
    &\hat{a}_{l}(x) \hat{\tilde{G}}^{+}_o(2k)|b;b\rangle =2\sqrt{2}\sum_{\alpha'_1,\alpha'_2}\int \int dk'_1 dk'_2~\frac{|\tilde{\Phi}_{1}(\alpha'_{2},k'_{2})\rangle \tilde{\phi}_{l}^{\alpha'_{1}}(x,k'_{1}) }{2k-(k'_1+k'_2)+i 0^{+}} \big(\tilde{\phi}_{b}^{\alpha'_{1}}(k'_{1})\tilde{\phi}_{b}^{\alpha'_{2}}(k'_{2})\big)^{*}.
    \end{align}
This gives us $\langle b;b|\hat{\tilde{G}}^{-}_o(2k) \hat{a}^\dagger_{l}(x)\hat{a}_{l}(x) \hat{\tilde{G}}^{+}_o(2k)|b;b\rangle
   =8\sum_{\alpha'}\int  dk' |\tilde{\phi}_{b}^{\alpha'}(k')\tilde{\mathbf{I}}_{l}(x,2k-k')|^{2}$
where $\alpha'=R,L$. We evaluate the non-vanishing integrals as follows.
\begin{align}
    \textbf{(a) Integration :}~~~&\sum_{\alpha'}\int  dk' |\tilde{\phi}_{b}^{\alpha'}(k')\tilde{\mathbf{I}}_{2}(x,2k-k')|^{2} \nonumber\\
    =&\frac{1}{2\pi} \int  dk' \frac{  \big(| g_L^*-iJ^*g_R^*|^2+| g_R^*-iJg_L^*|^2\big)|g_R-iJ^{*} g_L |^2/(|J|^2+1)^4}{({k}-\Omega-\Delta+i\Gamma/2)({k'}-\Omega-\Delta-i\Gamma/2)(E-{k'}-\Omega-\Delta+i\Gamma/2)(E-{k'}-\Omega-\Delta-i\Gamma/2)}\nonumber\\
    =&2 \frac{  (| g_L^*-iJ^*g_R^*|^2+| g_R^*-iJg_L^*|^2)|g_R-iJ^{*} g_L |^2}{ \Gamma(|J|^2+1)^4[({2k}-2\Omega-2\Delta)^2+\Gamma^2]}=\frac{2}{(|J|^2+1)^2} \frac{  |g_R-iJ^{*} g_L |^2}{ ({2k}-2\Omega-2\Delta)^2+\Gamma^2},
\end{align}
where we enclose the integration contour in the lower half of the complex plane (because $x>0$). In a similar manner, we get
\begin{align}
    \textbf{(b) Integration :}~~~&\sum_{\alpha'}\int  dk' |\tilde{\phi}_{b}^{\alpha'}(k')\tilde{\mathbf{I}}_{1}(x,2k-k')|^{2} =\frac{2}{(|J|^2+1)^2} \frac{  |g_L-iJ g_R |^2}{ ({2k}-2\Omega-2\Delta)^2+\Gamma^2},~~~\text{for}~x>0.~~~~~~~~~~~~~~~
\end{align}
After simplification, we yield the following two expressions for the purely interaction-induced nonlinear currents
\begin{align}
   &\delta \tilde{J}^{c}_{L\to R}(2;k)
   =\frac{4 |g_R-iJ^{*} g_L|^2}{(|J|^2+1)^2} |\tilde{\phi}_{b}^{R}(k)|^4,\\
   &\delta \tilde{J}^{c}_{R\to L}(2;k)
   =\frac{4 |g_L-iJ g_R|^2}{(|J|^2+1)^2} |\tilde{\phi}_{b}^{L}(k)|^4. 
\end{align}
Once again, we observe $\delta \tilde{J}^{s}_{L\to R}(2;k) \neq \delta \tilde{J}^{s}_{R\to L}(2;k) $ for the generic cases when $|\tilde{\phi}_{b}^{R}(k)|\neq |\tilde{\phi}_{b}^{L}(k)|$. Hence, the total nonlinear current is nonreciprocal,  $\delta \tilde{J}_{R\to L}(2;k)\neq \delta \tilde{J}_{R\to L}(2;k)$, for the generic cases.

Let us look into the special cases, (a) taking all  parameters $g_L,g_R,J$ to be reals with $J=\pm 1$, we see  $\delta \tilde{J}^{s}_{L\to R}(2;k) = \delta \tilde{J}^{s}_{R\to L}(2;k)$ while keeping $g_L \neq g_R$.
Next, (b) choosing the coupling parameters according to Eqs.~(\ref{Eq_gLTrans_Supple}-\ref{Eq_gRTrans_Supple}) with $J=-i$, we get $\delta \tilde{J}^{s}_{L\to R}(2;k)=\delta \tilde{J}^{s}_{R\to L}(2;k)$, because $|g_R+g_L|^2=|g_R-g_L|^2=2\Gamma$, and $|\tilde{\phi}_{b}^{R}(k)|^2=|\tilde{\phi}_{b}^{L}(k)|^2$. Therefore, for the special cases (a) and (b), one revives the reciprocity in the nonlinear parts of the 2-photon currents when  $|\tilde{\phi}_{b}^{R}(k)|=|\tilde{\phi}_{b}^{L}(k)|$.
We further see, if we choose $\theta=\Omega$ in the mapping relations in Eqs.~(\ref{Eq_gLTrans_Supple}-\ref{Eq_gRTrans_Supple}) with $J=-i$, we obtain 
\begin{align}
   \delta\tilde{J}_{R\to L}(2;k)= \delta\tilde{J}^{s}_{R\to L}(2;k)
   =\frac{\Gamma_{\Omega}^3 }{2(2\pi)^2\big[({k}-\Omega-\Delta_{\Omega})^2+\Gamma^2_{\Omega}/4\big]^2},
\end{align}
which is the nonlinear part of the 2-photon  transmission current in the giant-atom waveguide setup within the Markov approximation (see Eq.~\ref{MArkov_Gianttwophoton_intcurrent}).

\subsection{Elastic and inelastic components of the scattering matrix}
In this manuscript, we show a transition from the nonreciprocal to the reciprocal transport regimes in the generalized direct-coupled setup can be identified through the inelastic component of the scattering matrix for two correlated photons. To access both the elastic and inelastic components of the scattering matrix, we use the Lippmann–Schwinger formalism to calculate the outgoing states for all possible configurations of incident directions and momenta of two photons.

\subsubsection{Outgoing states}
As before, the scattered 2-photon state is denoted by
\begin{align}
|\tilde{\Psi}_2(\alpha,k_1;\beta,k_2)\rangle=&|\tilde{\Phi}_2(\alpha,k_1;\beta,k_2)\rangle- \frac{\hat{\tilde{G}}^{+}_o(E_{\mathbf k})|b;b\rangle}{\langle b;b|\hat{\tilde{G}}^{+}_o(E_{\mathbf k})|b;b\rangle}\langle b;b|\tilde{\Phi}_{2}(\alpha,k_1;\beta,k_2)\rangle.
\end{align}
where $\alpha,\beta=R,R$; $\alpha,\beta=R,L$; and $\alpha,\beta=L,L$.  One of the incident photons is moving towards $\alpha$ with momentum $k_1$, and the other one is moving towards $\beta$ with momentum $k_2$. Since the photons are injected in the product state, we take
\begin{align}
|\tilde{\Phi}_{2}(\alpha,k_1;\beta,k_2)\rangle=&\frac{1}{\sqrt{1+\delta_{k_1,k_2} \delta_{\alpha,\beta}}}|\tilde{\Phi}_{1}(\alpha,k_1)\rangle |\tilde{\Phi}_{1}(\beta,k_2)\rangle.
\end{align}
We take spatial projections of the scattered 2-photon state in the outgoing regions  $x_1,x_2>0$ of the channels $l=1,2$. The resulting wavefunctions are then analytically extended to all spatial regions to obtain the outgoing waves.
Here, $E_{\mathbf k}=k_1+k_2$ is the total energy of the incident photons. In the following we consider three cases for 
all possible configurations of incident directions.

\textbf{\textrm{I}.  Two incident right-moving photons}.--
The outgoing state in which both photons are scattered into channel-1 is
\begin{align}
    &\langle x_1,1;x_2,1|\tilde{\Psi}_2(R,k_1;R,k_2)\rangle=\tilde{r}^{R}_{k_1} \tilde{r}^{R}_{k_2}\frac{e^{i(k_1x_1+k_2x_2)}+e^{i(k_1x_2+k_2x_1)}}{2\pi \sqrt{2+2\delta_{k_1,k_2}}}+\bigg[\frac{(g_L-i J g_R )}{|J|^2+1}\bigg]^2 e^{i E_{\mathbf k} x_c}e^{i (E_{\mathbf k}-\Omega-\Delta+i\Gamma/2)|x|}\frac{\sqrt{2}\tilde{\phi}_{b}^{R}(k_1)\tilde{\phi}_{b}^{R}(k_2)}{\sqrt{1+\delta_{k_1,k_2}}}.\label{S133_twoPhoton}
\end{align}
The outgoing state in which one photon exits through channel-1 and the other through channel-2 is given by
\begin{align}
    &\langle x_1,1;x_2,2|\tilde{\Psi}_2(R,k_1;R,k_2)\rangle\nonumber\\
    =&\frac{\tilde{t}^{R}_{k_2} \tilde{r}^{R}_{k_1}e^{i(k_1x_1+k_2x_2)}+\tilde{t}^{R}_{k_1} \tilde{r}^{R}_{k_2}e^{i(k_1x_2+k_2x_1)}}{2\pi \sqrt{1+\delta_{k_1,k_2}}}+\sqrt{2} \bigg[\frac{(g_R-i J^{*} g_L )}{(|J|^2+1)} \frac{(g_L-i J g_R )}{(|J|^2+1)}\bigg] e^{i E_{\mathbf k} x_c}e^{i (E_{\mathbf k}/2-\Omega-\Delta+i\Gamma/2)|x|}\frac{\sqrt{2}\tilde{\phi}_{b}^{R}(k_1)\tilde{\phi}_{b}^{R}(k_2)}{\sqrt{1+\delta_{k_1,k_2}}}.\label{S134_twoPhoton}
\end{align}
The outgoing state in which both photons exit through channel-2
\begin{align}
    &\langle x_1,2;x_2,2|\tilde{\Psi}_2(R,k_1;R,k_2)\rangle
    =\tilde{t}^{R}_{k_1} \tilde{t}^{R}_{k_2}\frac{e^{i(k_1x_1+k_2x_2)}+e^{i(k_1x_2+k_2x_1)}}{2\pi \sqrt{2+2\delta_{k_1,k_2}}}+\bigg[\frac{(g_R-i J^* g_L )}{|J|^2+1}\bigg]^2 e^{i E_{\mathbf k} x_c}e^{i (E_{\mathbf k}/2-\Omega-\Delta+i\Gamma/2)|x|}\frac{\sqrt{2}\tilde{\phi}_{b}^{R}(k_1)\tilde{\phi}_{b}^{R}(k_2)}{\sqrt{1+\delta_{k_1,k_2}}}.\label{S135_twoPhoton}
\end{align}
\textbf{\textrm{II}.  One incoming photon from the left and another  from the right}.--
The outgoing state in which both photons are scattered into channel-1 is
\begin{align}
    &\langle x_1,1;x_2,1|\tilde{\Psi}_2(R,k_1;L,k_2)\rangle=\tilde{r}^{R}_{k_1} \tilde{t}^{L}_{k_2}\frac{e^{i(k_1x_1+k_2x_2)}+e^{i(k_1x_2+k_2x_1)}}{2\pi \sqrt{2}}+\sqrt{2}\bigg[\frac{(g_L-i J g_R )}{|J|^2+1}\bigg]^2 e^{i E_{\mathbf k} x_c}e^{i (E_{\mathbf k}/2-\Omega-\Delta+i\Gamma/2)|x|}\tilde{\phi}_{b}^{R}(k_1)\tilde{\phi}_{b}^{L}(k_2).\label{S138_twoPhoton}
\end{align}
The outgoing state in which one photon exits through channel-1 and the other through channel-2 is given by
\begin{align}
    &\langle x_1,1;x_2,2|\tilde{\Psi}_2(R,k_1;L,k_2)\rangle\nonumber\\
    =&\frac{\tilde{r}^{R}_{k_1} \tilde{r}^{L}_{k_2}e^{i(k_1x_1+k_2x_2)}+\tilde{t}^{R}_{k_1} \tilde{t}^{L}_{k_2}e^{i(k_1x_2+k_2x_1)}}{2\pi}+2 \bigg[\frac{(g_R-i J^{*} g_L )}{(|J|^2+1)} \frac{(g_L-i J g_R )}{(|J|^2+1)}\bigg] e^{i E_{\mathbf k} x_c}e^{i (E_{\mathbf k}/2-\Omega-\Delta+i\Gamma/2)|x|}\tilde{\phi}_{b}^{R}(k_1)\tilde{\phi}_{b}^{L}(k_2).\label{S139_twoPhoton}
\end{align}
The outgoing state in which both photons exit through channel-2
\begin{align}
    &\langle x_1,2;x_2,2|\tilde{\Psi}_2(R,k_1;L,k_2)\rangle= \tilde{t}^{R}_{k_1} \tilde{r}^{L}_{k_2}\frac{ e^{i(k_1x_1+k_2x_2)}+ e^{i(k_1x_2+k_2x_1)}}{2\pi \sqrt{2}}+\sqrt{2}\bigg[\frac{(g_R-i J^* g_L )}{|J|^2+1}\bigg]^2 e^{i E_{\mathbf k} x_c}e^{i (E_{\mathbf k}/2-\Omega-\Delta+i\Gamma/2)|x|}\tilde{\phi}_{b}^{R}(k_1)\tilde{\phi}_{b}^{L}(k_2).\label{S140_twoPhoton}
\end{align}
\textbf{\textrm{III}.  Two incident left-moving photons}.-- 
The outgoing state in which both photons are scattered into channel-1 is
\begin{align}
    &\langle x_1,1;x_2,1|\tilde{\Psi}_2(L,k_1;L,k_2)\rangle
    =\tilde{t}^{L}_{k_1} \tilde{t}^{L}_{k_2}\frac{e^{i(k_1x_1+k_2x_2)}+e^{i(k_1x_2+k_2x_1)}}{2\pi \sqrt{2+2\delta_{k_1,k_2}}}+\bigg[\frac{(g_L-i J g_R )}{|J|^2+1}\bigg]^2 e^{i E_{\mathbf k} x_c}e^{i (E_{\mathbf k}/2-\Omega-\Delta+i\Gamma/2)|x|}\frac{\sqrt{2}\tilde{\phi}_{b}^{L}(k_1)\tilde{\phi}_{b}^{L}(k_2)}{\sqrt{1+\delta_{k_1,k_2}}}.\label{S143_twoPhoton}
\end{align}
The outgoing state in which one photon exits through channel-1 and the other through channel-2 is given by
\begin{align}
    &\langle x_1,1;x_2,2|\tilde{\Psi}_2(L,k_1;L,k_2)\rangle\nonumber\\
    =&\frac{\tilde{t}^{L}_{k_1} \tilde{r}^{L}_{k_2}e^{i(k_1x_1+k_2x_2)}+\tilde{t}^{L}_{k_2} \tilde{r}^{L}_{k_1}e^{i(k_1x_2+k_2x_1)}}{2\pi \sqrt{1+\delta_{k_1,k_2}}}+\sqrt{2} \bigg[\frac{(g_R-i J^{*} g_L )}{(|J|^2+1)} \frac{(g_L-i J g_R )}{(|J|^2+1)}\bigg] e^{i E_{\mathbf k} x_c}e^{i (E_{\mathbf k}/2-\Omega-\Delta+i\Gamma/2)|x|}\frac{\sqrt{2}\tilde{\phi}_{b}^{L}(k_1)\tilde{\phi}_{b}^{L}(k_2)}{\sqrt{1+\delta_{k_1,k_2}}}.\label{S144_twoPhoton}
\end{align}
The outgoing state in which both photons exit through channel-2
\begin{align}
    &\langle x_1,2;x_2,2|\tilde{\Psi}_2(L,k_1;L,k_2)\rangle=\tilde{r}^{L}_{k_1} \tilde{r}^{L}_{k_2}\frac{e^{i(k_1x_1+k_2x_2)}+e^{i(k_1x_2+k_2x_1)}}{2\pi \sqrt{2+2\delta_{k_1,k_2}}}+\bigg[\frac{(g_R-i J^* g_L )}{|J|^2+1}\bigg]^2 e^{i E_{\mathbf k} x_c}e^{i (E_{\mathbf k}/2-\Omega-\Delta+i\Gamma/2)|x|}\frac{\sqrt{2}\tilde{\phi}_{b}^{L}(k_1)\tilde{\phi}_{b}^{L}(k_2)}{\sqrt{1+\delta_{k_1,k_2}}}.\label{S145_twoPhoton}
\end{align}
From all the outgoing states, we construct the elastic and inelastic components of the scattering $\hat{\mathbf{S}}$-matrix.

\subsubsection{Elastic and inelastic components}
In the manuscript, we mention that the scattering $\mathbf{S}$-matrix generates a mapping between two plane-wave states. 
In the 2-photon sector, such a mapping captures the essence of elastic and inelastic scattering as follows \cite{ShenFanPRL2007}: 
\begin{align}
    \langle S_{q_1,q_2}(m,m')|\hat{\mathbf{S}}|S_{k_1,k_2}(l,l')\rangle=\Big[\mathbf{M}_{m,m'}^{l,l'}(k_1,k_2)\delta(q_1-k_1) \delta(q_2-k_2)+(k_1 \leftrightarrow k_2)\Big]+ \mathbf{N}_{m,m'}^{l,l'}(\mathbf q;\mathbf k)\delta(E_{\mathbf q}-E_{\mathbf k}), \label{S152_two_photon}
\end{align}
where the plane-wave states in the 2-photon sector are defined as
\begin{align}
&|S_{k_1,k_2}(l,l')\rangle=\int \int dx_1 dx_2 \frac{e^{i(k_1x_1+k_2x_2)}}{2\pi }\hat{a}^\dagger_{l}(x_1) \hat{a}^\dagger_{l'}(x_2)|\varphi\rangle,~~~\text{for channels}~~l,l'=1,2.
\end{align}
These states satisfy the following identity resolution in the waveguides' 2-photon  subspace:
\begin{align}
    \mathds{1}_2 
    &=\frac{1}{2}\sum_{l=1,2}\int_{-\infty}^{\infty} \int_{-\infty}^{\infty}  dk_1 dk_2 |S_{k_1,k_2}(l,l)\rangle \langle S_{k_1,k_2}(l,l)|+\int_{-\infty}^{\infty} \int_{-\infty}^{\infty}  dk_1 dk_2 |S_{k_1,k_2}(1,2)\rangle \langle S_{k_1,k_2}(1,2)|.
\end{align}
The first two terms in the mapping relation Eq.~\ref{S152_two_photon} describe elastic scattering via direct and exchange processes, while the last term corresponds to inelastic scattering, in which the total energy is conserved but the individual photon momentum may change. Here, $k_1,k_2$ are the momenta of the incident photons, whereas  $q_1,q_2$ are the momenta of the scattered photons.
For an incident plane-wave state $|X_{\rm in}\rangle$, the corresponding outgoing state is obtained by the action of the the $\hat{\mathbf{S}}$-matrix as follows:  $|X_{\rm out}\rangle=\hat{\mathbf{S}}|X_{\rm in}\rangle$. We consider the following input states corresponding to the different possible directions of the incident photons:

\textbf{Case \textrm{I}.  Two incident right-moving photons}. $|X_{\text{in}}\rangle=  |S_{k_1,k_2}(1,1)\rangle/\sqrt{1+\delta_{k_1,k_2}}$;

\textbf{Case \textrm{II}.  One incoming photon from the left and another  from the right}. $|X_{\text{in}}\rangle=  |S_{k_1,k_2}(1,2)\rangle$;

\textbf{Case \textrm{III}.  Two incident left-moving photons}. $|X_{\text{in}}\rangle=  |S_{k_1,k_2}(2,2)\rangle/\sqrt{1+\delta_{k_1,k_2}}$.

\noindent
We then evaluate the projections$\langle x_1, l;x_2,l'|X_{\text{out}}\rangle$ for channels $l,l'=1,2$ and compare them with Eqs.~\ref{S133_twoPhoton}--\ref{S135_twoPhoton} for the Case~\textbf{\textrm{I}}, Eqs.~\ref{S138_twoPhoton}--\ref{S140_twoPhoton} for the Case~\textbf{\textrm{II}}, and Eqs.~\ref{S143_twoPhoton}--\ref{S145_twoPhoton} for the Case~\textbf{\textrm{III}}. Such a comparison unambiguously determines the elements $\mathbf{M}_{m,m'}^{l,l'}(k_1,k_2)$ and $\mathbf{N}_{m,m'}^{l,l'}(\mathbf q;\mathbf k)$.  We can arrange the elements in  $3\times 3$ matrix representation, where the index pairs $(m,m')$ and $(l,l')$ runs over the set $((1,1), (1,2), (2,2))$. The subscript $(m,m')$ denotes the rows and the superscript  $(l,l')$ denotes the columns. We finally get
\begin{align}
  \mathbf{M}(k_1,k_2)=  \begin{bmatrix}
         \tilde{r}^R_{k_1}  \tilde{r}^R_{k_2}  &  \tilde{r}^R_{k_1}  \tilde{t}^L_{k_2}& \tilde{t}^L_{k_1}  \tilde{t}^L_{k_2} \\
         \tilde{t}^R_{k_1}  \tilde{r}^R_{k_2} & \tilde{t}^R_{k_1}  \tilde{t}^L_{k_2} & \tilde{r}^L_{k_1}  \tilde{t}^L_{k_2} \\
         \tilde{t}^R_{k_1}  \tilde{t}^R_{k_2} &  \tilde{t}^R_{k_1}  \tilde{r}^L_{k_2} &\tilde{r}^L_{k_1}  \tilde{r}^L_{k_2}
    \end{bmatrix}, \label{M_matrix_supple_2photon}
\end{align}
and 

\begin{align}
  &\mathbf{N}(\mathbf q;\mathbf k)=\begin{bmatrix}
        (g_L-i J g_R)^2(g^*_L-i J^*g^*_R)^2  &  (g_L-i J g_R)^2(g^*_L-i J^*g^*_R)(g^*_R-i J g^*_L)& (g_L-i J g_R)^2(g^*_R-i J g^*_L)^2 \\
        &&\\
         \splitfrac{(g_L-i J g_R) (g_R-i J^*g_L)}{\times(g^*_L-i J^*g^*_R)^2} & \splitfrac{(g_L-i J g_R) (g_R-i J^*g_L)}{\times(g^*_L-i J^*g^*_R)(g^*_R-i J g^*_L)} & \splitfrac{(g_L-i J g_R) (g_R-i J^*g_L)}{\times(g^*_R-i J g^*_L)^2} \\
         &&\\
         (g_R-i J^*g_L)^2(g^*_L-i J^*g^*_R)^2 &  (g_R-i J^*g_L)^2(g^*_L-i J^*g^*_R)(g^*_R-i J g^*_L) & (g_R-i J^*g_L)^2(g^*_R-i J g^*_L)^2
    \end{bmatrix}h(\mathbf q;\mathbf k),\label{N_matrix_supple_2photon}
\end{align}
where 
\begin{align}
     h(\mathbf q;\mathbf k)=-\frac{16i}{\pi} \frac{(E_{\mathbf k}-2\Omega-2\Delta+i\Gamma)/(|J|^2+1)^4}{[4\delta_{\mathbf k}^2-(E_{\mathbf k}-2\Omega-2\Delta+i\Gamma)^2][4\delta_{\mathbf q}^2-(E_{\mathbf k}-2\Omega-2\Delta+i\Gamma)^2]}. \label{N2_matrix_supple_2photon}
\end{align}

We define the relative momentum for the incident photons and scattered photons using $\delta_{\mathbf k}=(k_1-k_2)/2$ and $\delta_{\mathbf q}=(q_1-q_2)/2$. 
In the above calculations, we use the following relations:
\begin{align}
   1.& \langle S_{q_1,q_2}(l,l)|S_{k_1,k_2}(l,l)\rangle =\delta(E_{\mathbf q}-E_{\mathbf k})\big[\delta(\delta_{\mathbf q}-\delta_{\mathbf k})+\delta(\delta_{\mathbf q}+\delta_{\mathbf k})\big]=\delta(q_1-k_1)\delta(q_2-k_2)+\delta(q_1-k_2)\delta(q_2-k_1),\\
    2.&  \langle S_{q_1,q_2}(1,2)|S_{k_1,k_2}(1,2)\rangle=\delta(q_1-k_1)\delta(q_2-k_2),\\
    3.&\langle x_1,l; x_2,l|S_{q_1,q_2}(l,l)\rangle=\frac{e^{i(q_1x_1+q_2x_2)}+e^{i(q_1x_2+q_2x_1)}}{2\pi \sqrt{2}}= \frac{\sqrt{2}}{2\pi} e^{iE_{\mathbf q} x_c} \cos{\delta_{\mathbf q} x},\\
    4.&\langle x_1,1; x_2,2|S_{q_1,q_2}(1,2)\rangle=\frac{e^{i(q_1x_1+q_2x_2)}}{2\pi }= \frac{1}{2\pi} e^{iE_{\mathbf q} x_c} e^{i\delta_{\mathbf q} x},\\
    5.&\int_{-\infty}^{\infty} d\delta_{\mathbf q} \frac{\cos{\delta_{\mathbf q}x}}{[4\delta_{\mathbf q}^2-(E_{\mathbf k}-2\Omega-2\Delta+i\Gamma)^2]}
    =\frac{1}{4}(-2\pi i) \frac{e^{i(E_{\mathbf k}-2\Omega-2\Delta+i\Gamma)|x|/2}}{(E_{\mathbf k}-2\Omega-2\Delta+i\Gamma)}~~~(\because \Gamma>0).
\end{align}

\bigskip
\hrule 
\bigskip

\end{document}